\documentclass[12pt,preprint]{aastex}

\newcommand{\MSUNYR}{M_\odot/yrs}

\begin{document}

\title{Resolving the CO Snow Line in the Disk around HD~163296}

\author{Chunhua~Qi\altaffilmark{1}, 
Paola D'Alessio\altaffilmark{2}, 
Karin I. \"Oberg\altaffilmark{1,3}, 
David J. Wilner\altaffilmark{1},
A. Meredith Hughes\altaffilmark{1,4,5},
Sean M. Andrews\altaffilmark{1}, 
Sandra Ayala\altaffilmark{2} } 

\altaffiltext{1}{Harvard--Smithsonian Center for Astrophysics, 
60 Garden Street, MS 42, Cambridge, MA 02138, USA; 
cqi, koberg, dwilner, mhughes, sandrews@cfa.harvard.edu.} 

\altaffiltext{2}{Centro de Radioastronom\'{i}a y Astrof\'{i}sica, 
Universidad Nacional Aut\'{o}noma de M\'{e}xico, 58089 Morelia, 
Michoac\'{a}n, M\'{e}xico; p.dalessio@crya.unam.mx, sayala2001@gmail.com}

\altaffiltext{3}{Hubble Fellow}

\altaffiltext{4}{Department of Astronomy, University of California at
  Berkeley, 601 Campbell Hall, Berkeley, CA 94720, USA}

\altaffiltext{5}{Miller Fellow}

\begin{abstract}
We report Submillimeter Array (SMA) observations of CO (J=2--1, 3--2
and 6--5) and its isotopologues ($^{13}$CO J=2--1, C$^{18}$O J=2--1
and C$^{17}$O J=3--2) in the disk around 
the Herbig~Ae star HD~163296 at $\sim2''$ (250 AU) resolution, and
interpret these data in the framework of a model that constrains the
radial and vertical location of the line emission regions.
First, we develop a physically
self-consistent accretion disk model with an exponentially tapered
edge that matches the spectral energy distribution and spatially
resolved millimeter dust continuum emission. Then, we 
refine the vertical structure of the model using wide range of
excitation conditions sampled by the CO lines, in particular the
rarely observed J=6--5 transition. By fitting
$^{13}$CO data in this structure, we further constrain the
vertical distribution of CO to lie between a lower
boundary below which CO freezes out
onto dust grains (T $\lesssim 19$ K) and an upper boundary above which CO
can be photodissociated (the hydrogen column density from the disk
surface is $\lesssim10^{21}$ cm$^{-2}$). The freeze-out at 19 K leads
to a significant drop in the gas-phase CO column density beyond a radius of
$\sim$155~AU,  a ``CO snow line'' that we
directly resolve. By fitting the abundances of all CO isotopologues,
we derive isotopic 
ratios of $^{12}$C/$^{13}$C, $^{16}$O/$^{18}$O and $^{18}$O/$^{17}$O  
that are consistent with quiescent interstellar gas-phase values.  
This detailed model of the HD~163296 disk demonstrates the
potential of a staged, parametric technique for constructing unified
gas and dust structure models and constraining the distribution of
molecular abundances using resolved multi-transition, multi-isotope
observations.

\end{abstract}

\keywords {circumstellar matter ---techniques: interferometric
  ---planetary systems: protoplanetary disks ---stars:
  individual(HD~163296) ---ISM: abundances ---radio lines: stars} 

\section{Introduction}
The disks around pre-main sequence stars are the reservoirs of raw material 
that represent the initial conditions for the formation of planetary systems.  
The spatial distribution of mass in these disks is a fundamental property, as 
it sets a constraint on the contents and orbital architectures of planetary 
systems.  In that sense, measurements of disk densities and temperatures can 
provide strong, albeit complex, constraints on planet-forming scenarios.  
Because the dominant constituent of these disks is believed to be cold H$_2$ 
gas that is generally unobservable, our knowledge of many disk properties 
(including mass) relies on the observation and interpretation of minor 
constituents. Dust has naturally been the focus of the most work on disk 
structure and evolution, as it dominates the opacity and emits bright
continuum  radiation \citep[e.g.,][]{andrews09,andrews10}. 
However, converting those continuum surface brightnesses to disk (gas) 
densities is challenging due to the large uncertainties in the assumed dust 
properties (composition, size distribution, etc.) and the dust abundance 
relative to the gas, which may deviate substantially from the canonical 
interstellar value (0.01) and vary spatially in the disk.
In principle, spectral line emission from trace gas species can provide 
complementary constraints on the disk structure, but the interpretation of the 
observations can also be difficult.  These emission lines depend on the gas 
abundances that are set by chemical and physical processes, as well as the 
excitation conditions set by the local densities, temperatures, and incident 
radiation fields.  With a dataset of sufficient quality, these
complexities can be leveraged into powerful probes of otherwise
inaccessible disk characteristics \citep[e.g.,][]{kamp_t10}.

The chemical property of CO, the most abundant molecule after H$_2$,
is thought to be well understood in disks. The spatial distribution of
gas-phase CO is expected to be shaped by photodissociation both near
the star and high in the disk atmosphere, as well as by a depletion
process when it freezes onto dust grains at low temperatures (below
$\sim$20~K) at large radii and deep in the disk interior
\citep[e.g.,][]{aikawa_m96,aikawa_n06,gorti_h08}. The peak abundance
of CO is also predicted to be almost independent of radius, around
10$^{-4}$, the typical value in molecular clouds (\citealp{aikawa_n06}). 
However, observational studies of CO emission from disks have yielded
puzzling results.  
In general, the CO in disks is found to be under-abundant by factors
of 10--100 compared to molecular clouds, a deficit usually attributed
to CO depletion in cold, dense gas (\citealp{dutrey_g94, dutrey_g96,
  dutrey_g97, vanzadelhoff_v01, thi_v04}). 
However, multi-transition and multi-isotope CO studies 
report a wide range of relative depletions of CO and its
isotopologues.  For example, \citet{dartois_d03} finds that 
all of the CO isotopologues are depleted by an identical uniform
factor of $\sim$10, but \citet{pietu_d07} finds in a small sample of
disks that the $^{12}$CO/$^{13}$CO ratios in the outer part of disks 
are much lower than the standard $^{12}$C/$^{13}$C ratio in the solar
neighborhood: these low values were attributed to significant carbon
fractionation. At the same time, both of these studies suggested 
that the outer radius derived from $^{13}$CO is smaller than the one
derived from $^{12}$CO indicating their $^{12}$CO/$^{13}$CO ratios are
abnormally large at the outer disk edge of their models, which
leads to speculations on the importance of selective
photo-dissociation due to self-shielding in the photolysis of gaseous
CO (\citealp{dutrey_g07}). They also find that a large amount of CO
remains gaseous at temperatures as low as 10 K. Possible explanations
for the presence of the inferred cold CO include vertical  
mixing (\citealp{aikawa07}) and photodesorption off dust grains 
(\citealp{hersant_w09}). All above depict a very complex picture of
the distribution of CO gas in the disks, which remains to be explained by
detailed chemical models combined with realistic physical disk
structures. However, it is unclear, how much of these variations
depend on the individual sources, and how much, if any, depends on the
modeling approach. 

A robust methodology for interpreting CO observations is a high priority for 
our efforts to constrain the total gas distribution in disks, and to provide a 
basis for interpreting observations of more complex molecules.  In this study, 
we develop a modeling framework that is able to account for observations of 
both the dust continuum and CO line emission from a protoplanetary disk.  
Specifically, we combine a previously-established dust emission modeling 
formalism with multiple transitions, spatially resolved CO line data (CO J=2--1, 
3--2 and 6--5, $^{13}$CO J=2--1, C$^{18}$O J=2--1 and C$^{17}$O J=3--2) to
construct a simplified and internally self-consistent model of the
disk around the Herbig  Ae star HD~163296.  At a distance of $\sim$122
pc (\citealp{perryman_l97}), this 2.3 M$_{\odot}$ star (spectral type
A1; age $\sim$4 Myr) harbors a large disk with strong millimeter
continuum and molecular line emission
(\citealp{qi01,thi_v04,natta_t04}). It does not seem to be associated
with any known star-formation region, dark clouds, or reflection
nebulae. The disk around HD~163296 exhibits a scattered light pattern
extending out to a radius of $\sim$500 AU \citep{grady_d00},
perpendicular to a bipolar microjet traced by a chain of HH knots
visible in coronagraphic observations
(\citealp{devine_g00,grady_d00,wassell_g06}).  Millimeter and
submillimeter interferometric observations have explored this large
disk and its Keplerian velocity field in low-J CO lines
(\citealp{mannings_s97, isella_t07}). The large disk size and strong
molecular emission make this system an excellent target for resolved
millimeter observations and detailed modeling. 

This paper is organized as follows. We describe the CO and mm
continuum data in \S 2,
and present the main observational results in \S 3. Our extensive
modeling effort is detailed in \S 4. We discuss the modeling results
and compare with previous work on this subject in \S 5, and provide a
brief summary in \S 6.   

\section{Observations}

Observations of HD~163296 (RA: 17$^{\rm h}$56$^{\rm m}$21\fs279, DEC: 
$-$21\degr57$'$22\farcs09; J2000.0) were conducted between 2005 August and 
2010 September using the 8-antenna SMA\footnote{The Submillimeter
  Array is a joint project between the Smithsonian Astrophysical
  Observatory and the Academia Sinica Institute of Astronomy and
  Astrophysics, and is funded by the Smithsonian Institution and the
  Academia Sinica.} 
interferometer located atop Mauna Kea,  
Hawaii. Table 1 provides a general summary of the observational
parameters. 

For the observations of the CO 6--5 line, we used the correlator settings 
adopted by \citet{qi_w06} in their similar study of the TW Hya disk.  
At 690 GHz, there is no nearby quasar bright enough to use for phase
referencing with the SMA.  However, Callisto was located only 10\degr\
away from HD~163296 during our observation on 2007 March 20, and was
monitored every 20 minutes for use in the gain and absolute flux
calibration.  At that time, Callisto had a diameter of $1\farcs3$ and
a zero-spacing flux density of 48.1 Jy at 690 GHz. Based on the
uncertainties of the Callisto emission model, we estimate a 10\%\
systematic uncertainty in this adopted flux scale. The 690 GHz
continuum emission from HD~163296 was sufficiently bright (with a flux
density of 7.5 Jy) that we could correct the gain response of the SMA
with a single phase-only self-calibration iteration.

The observations of the CO 2--1 line took place on 2010 May 17 in the
compact configuration and 2010 September 14 in the extended configuration.
The correlator was configured to include 2048 channels in the 104 MHz
segment of the correlator centered on the rest frequency of the line. The
remaining 1.3 GHz of the correlator bandwidth in both sidebands were
configured with a uniform spectral resolution of 256 channels in each 104
MHz correlator segment to achieve maximum continuum sensitivity. The
weather was good on both nights, with the 225 GHz opacity just below 0.1
and stable atmospheric phase.  The nearby quasar 1733-130 (10.5 degrees
from the source) was used to calibrate the atmospheric and instrumental
phase, and the solution was checked using the quasar 1911-201, which was
included for a short period during each loop between source and calibrator.
Uranus and Neptune were used as flux calibrators on September 14, yielding
a 1733-130 flux of 2.71 Jy; Callisto was used to calibrate the May 17 flux,
yielding a 1733-130 flux of 2.04 Jy.

The observations of $^{13}$CO 2--1 and C$^{18}$O 2--1 were carried out
simultaneously on 2010 May 15 in the compact configuration and 2010
September 11 in the extended configuration. The correlator configuration
was identical to the CO 2--1 setup, except that the high resolution
correlator chunk was centered on the rest frequency of $^{13}$CO 2--1 with
C$^{18}$O 2--1 in a nearby low-resolution chunk.  The remaining 1.2 GHz were
devoted to uniform spectral resolution continuum observations.  The weather
was excellent on both nights with stable phase and 225 GHz opacity varying
between 0.04 (primarily during the extended track) and 0.06.  The same
calibrators as for the CO(2-1) observations were used; the derived flux of
1733-130 was 2.35 on September 11 and 2.04 Jy on May 15.

The observation of the C$^{17}$O 3--2 line took place on 2005 August 23 in
the compact configuration. The correlator was configured with a
uniform spectral resolution of 128 channels over 104 MHz, which
provided 0.8 MHz frequency resolution. The nearby quasars 1833-210 and
1921-293 were used as the calibrators. Uranus was used as flux
calibrator and the derived fluxes of 1833-210 and 1921-293 were 0.57
and 3.76 Jy, respectively.  

The observations of the CO 3--2 observations were summarized in
\citet{hughes_w11}. All of the calibration was 
performed using the MIR software 
package\footnote{http://www.cfa.harvard.edu/$\sim$cqi/mircook.html}.
Images of the continuum and the spectral lines were generated and
CLEANed using standard techniques in the MIRIAD software package.  

\section{Results}

\begin{figure}[htbp]
\centering
\includegraphics[width=6.5in,angle=90]{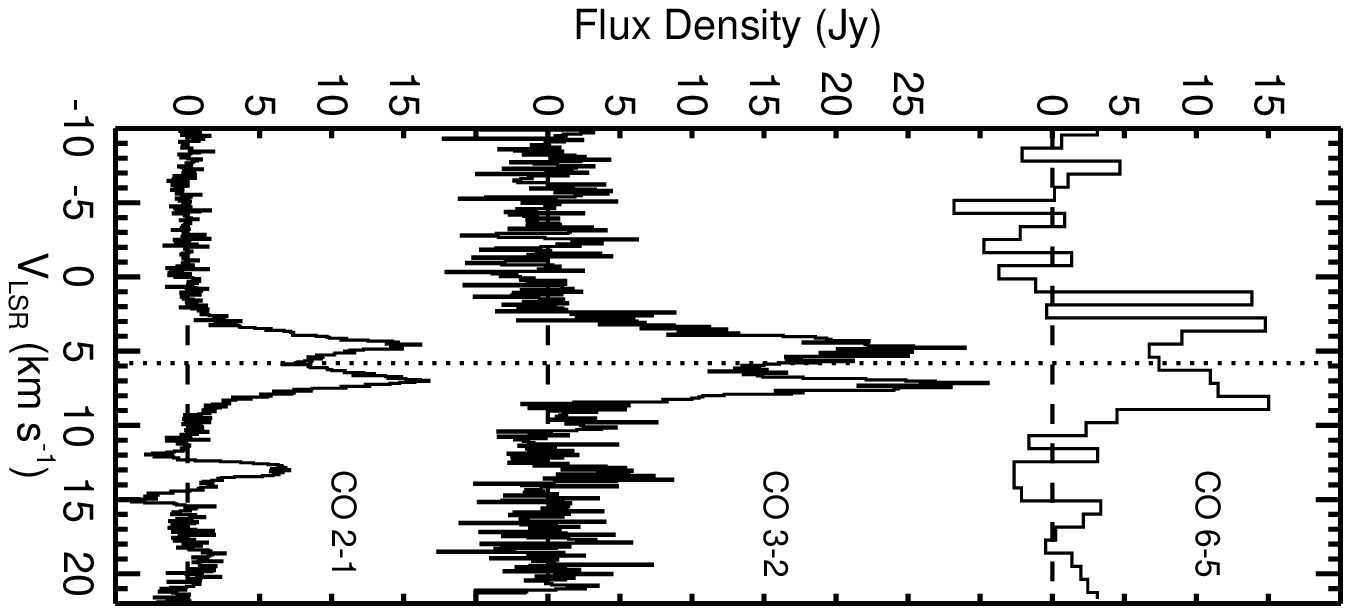}
\includegraphics[width=6.5in,angle=90]{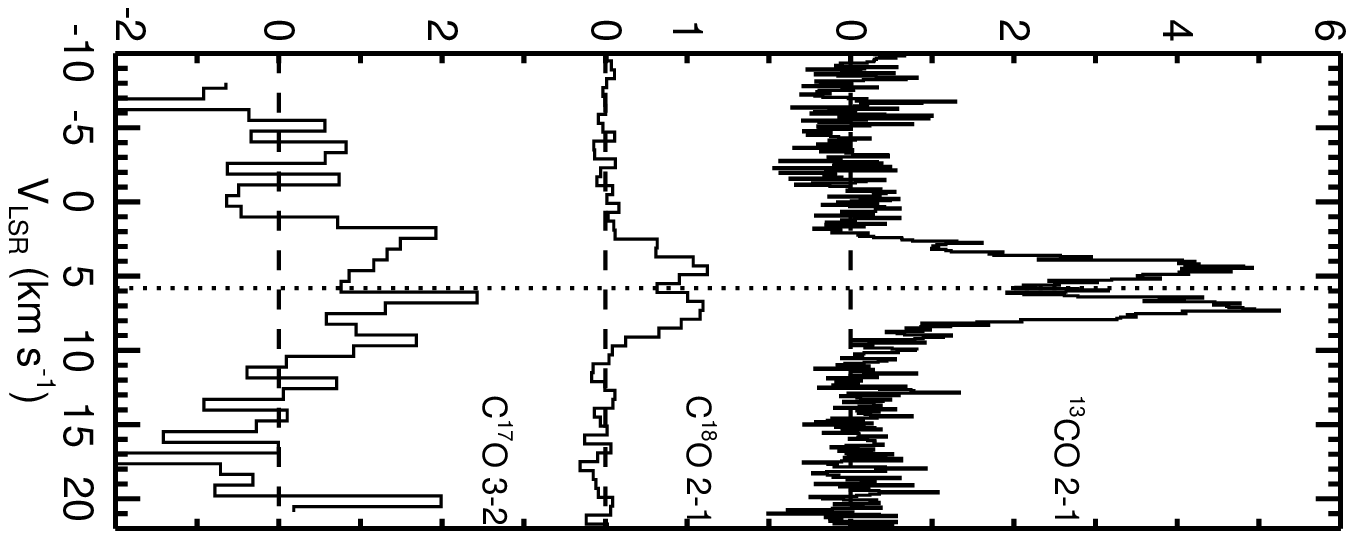}
\caption{ Spatially integrated spectra at the
peak continuum (stellar) position of HD~163296. The fluxes are
averaged over the emission areas (4$''$ box for CO 6--5; 10$''$ box
for other lines).     
The vertical dotted line indicates the position of the fitted V$_{LSR}$.
 \label{fig:specdata}}
\end{figure}

\begin{figure}[htbp]
\epsscale{1.0}
\plotone{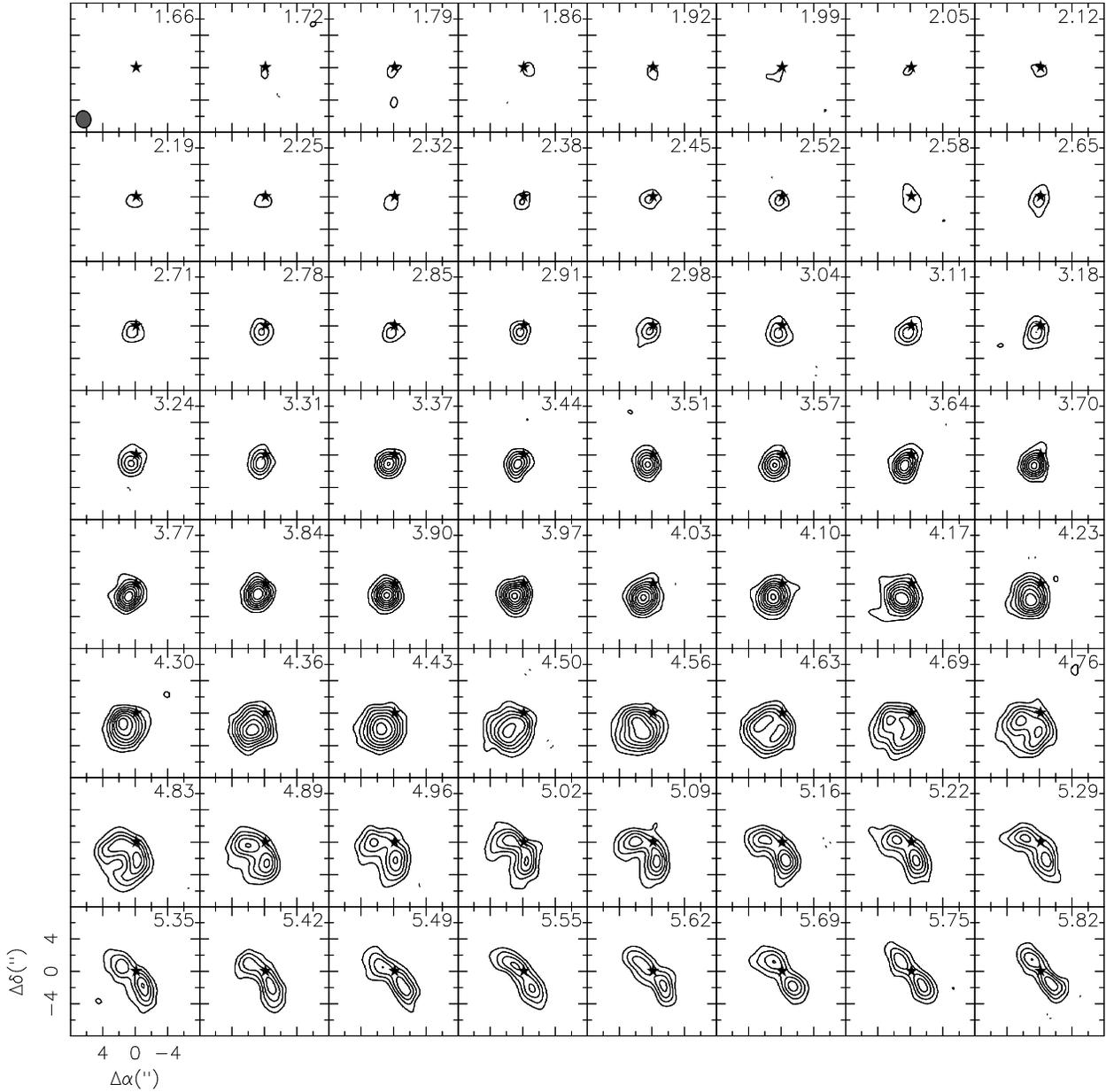}
\figcaption{
Channel maps of the CO J=2--1 line emission from the disk
  around HD~163296. The LSR velocity is indicated in the upper right
  of each channel, while the synthesized beam size and orientation
  ($2\farcs1 \times 1\farcs8$ at a position angle of 10.5$^{\circ}$)
  is indicated in the upper left 
  panel. The contours are 0.18 Jy Beam$^{-1}$ (1$\sigma$) $\times
  [3,6,9,12,15,18,21,24,27]$. The star symbol indicates the
  disk center.
 \label{fig:co21}}
\end{figure}

\begin{figure}[htbp]
\epsscale{1.0}
\plotone{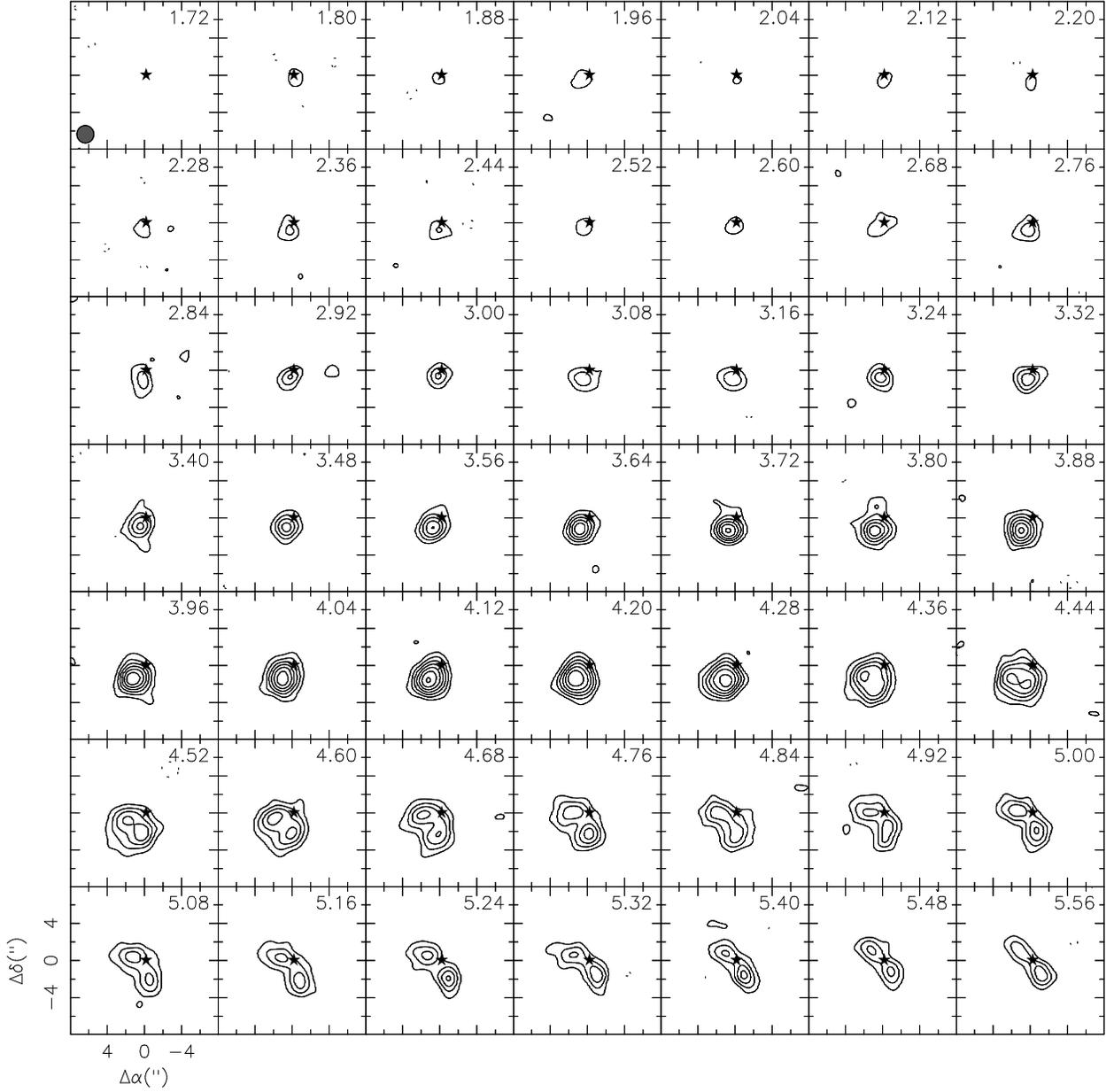}
\figcaption{
 Same as Figure~\ref{fig:co21}, but for $^{13}$CO 2--1. The
  beam size is $1\farcs9 \times 1\farcs8$ at a position angle of
  5.5$^{\circ}$ and the contours are 0.08 Jy Beam$^{-1}$ (1$\sigma$) $\times
  [3,6,9,12,15,18,21,24]$.
 \label{fig:13co21}}
\end{figure}

\begin{figure}[htbp]
\centering
\includegraphics[width=4.5in]{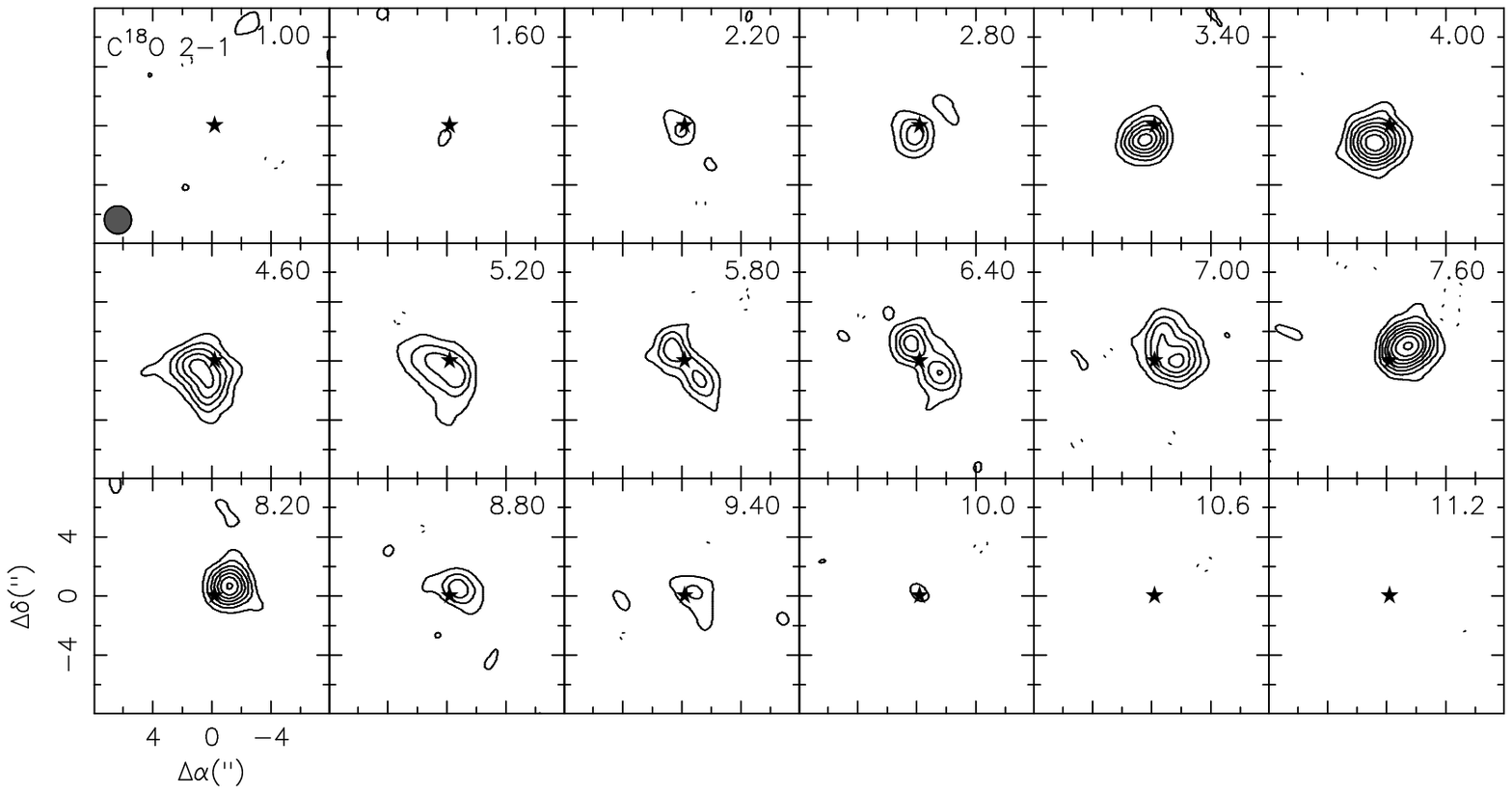}
\vskip 5mm
\includegraphics[width=4.5in]{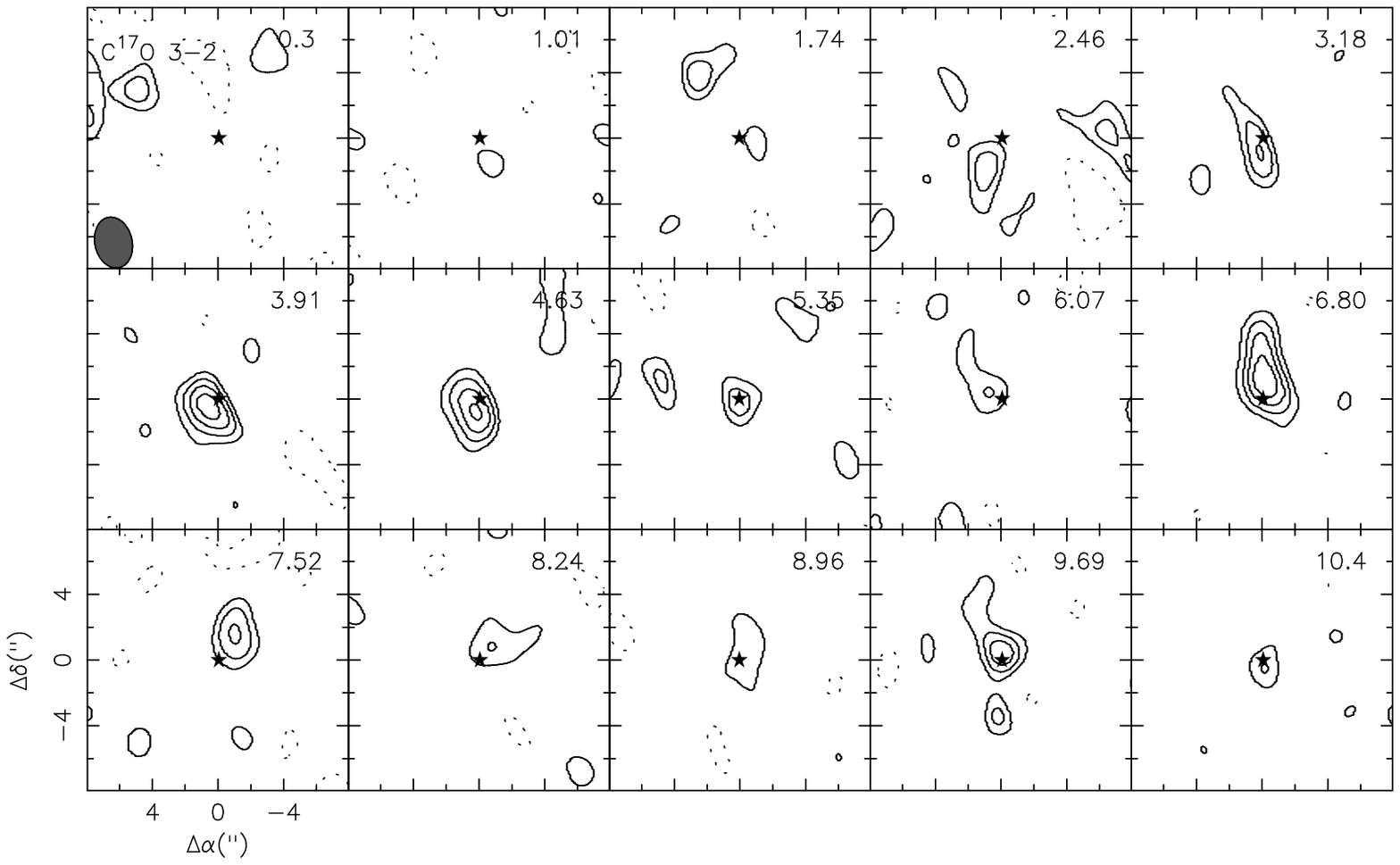}
\vskip 5mm
\includegraphics[width=4.5in]{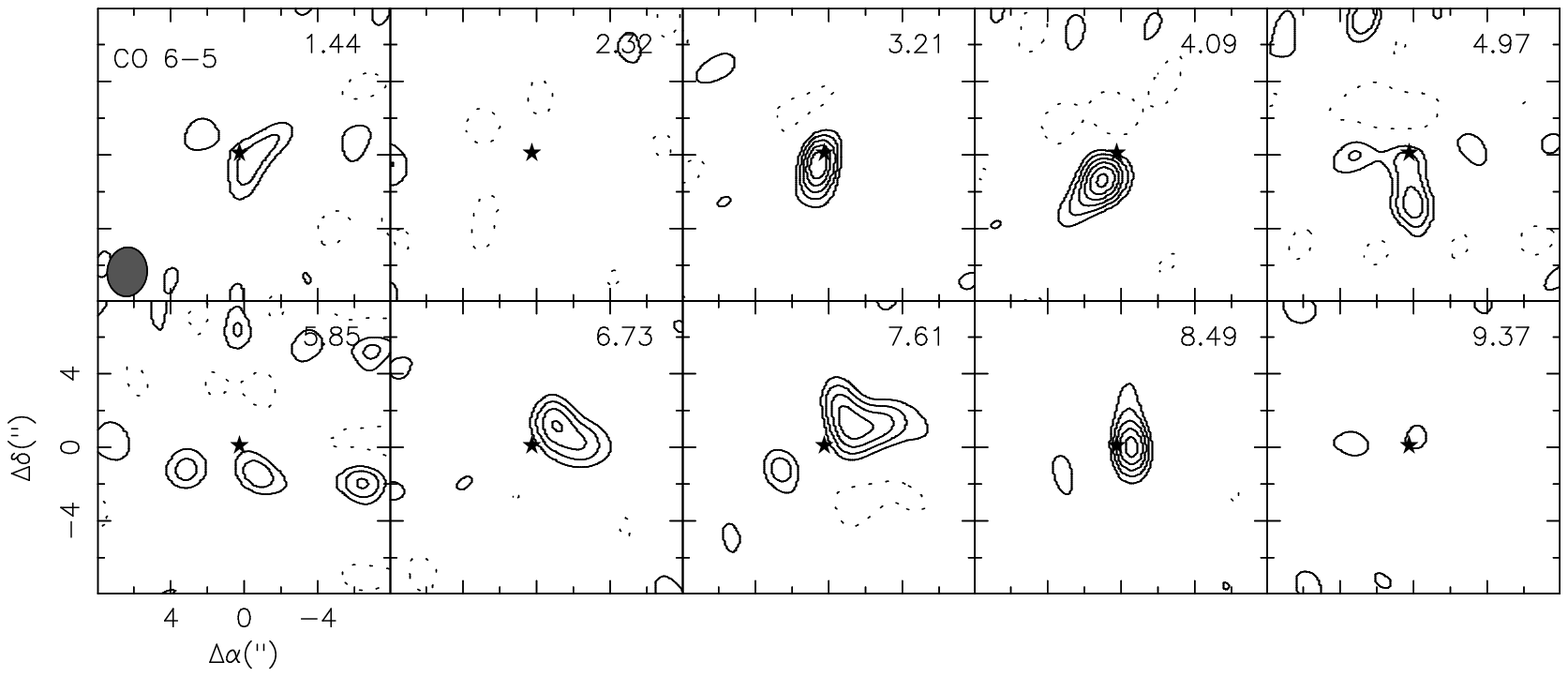}
\caption{ Same as Figure~\ref{fig:co21}, but for C$^{18}$O 2--1, C$^{17}$O
  3--2 and CO 6--5. For C$^{18}$O 2--1, the beam size is $1\farcs9
  \times 1\farcs8$ at a position angle of 
  5.5$^{\circ}$ and the contours are 0.025 Jy Beam$^{-1}$ (1$\sigma$) $\times
  [3,6,9,12,15,18,21,24]$; for C$^{17}$O 3--2, the beam size is $3\farcs1
  \times 2\farcs3$ at a position angle of 
  12.8$^{\circ}$ and the contours are 0.2 Jy Beam$^{-1}$ (1$\sigma$) $\times
  [2,3,4,5]$; for CO 6--5, the beam size is $2\farcs7
  \times 2\farcs2$ at a position angle of 
  -5.6$^{\circ}$ and the contours are 3.0 Jy Beam$^{-1}$ (1$\sigma$) $\times
  [2,3,4,5,6,7]$. 
 \label{fig:c18o21}}
\end{figure}

Figure~\ref{fig:specdata} shows the disk-averaged spectra of the CO J=2--1, 
3--2 and 6--5 lines, along with the J=2--1 lines of $^{13}$CO and
C$^{18}$O and the J=3--2 line of C$^{17}$O 3--2. These spectra were
extracted from the SMA channel maps in 10$''$ square boxes centered on
HD~163296, except for the CO 6--5 line. That fainter spectrum was
produced from a 4$''$-wide box chosen to cover the area with
statistically significant emission. Each emission line shows the
double-peaked velocity profile characteristic of disk rotation. The  
low-J CO lines have an additional feature in their spectra at $V_{\rm
  LSR}$ =  13 km s$^{-1}$, which is distinctly offset from the disk
kinematically and much more obvious in previous single-dish spectra
\citep{thi_v01, thi_v04}. That feature is apparently spatially
extended, and likely is associated with  
foreground and/or background molecular clouds. Table~2 lists the integrated 
intensities of the observed lines and the continuum flux densities.
Note that the signal-to-noise ratios range 
from over 200 for the  
CO 2--1 line to $\sim$10 for the CO 6--5 and C$^{17}$O 3--2
lines. Figure~\ref{fig:co21},~\ref{fig:13co21} and~\ref{fig:c18o21}
show the channel maps of all the 
transitions except for CO 3--2 which were presented in
\citet{hughes_w11}. 

\section{Analysis}

As a first step in developing a framework for understanding the chemical 
structures of protoplanetary disks, we aim to interpret this suite of 
observations in the context of an internally self-consistent model for the 
physical conditions of the gas and dust in the HD~163296 disk.  Our modeling 
analysis consists of three distinct stages.  First, we construct a template 
two-dimensional model for the disk structure based on observations of
the dust, using the broadband spectral energy distribution (SED) and
the resolved  
millimeter continuum emission as diagnostics (\S 4.1).  Second, we refine the 
vertical temperature distribution of that template structure based on
the spatially resolved line ratios of the optically thick CO
transitions (\S 4.2).  And third, we constrain 
the radial and vertical CO abundance pattern using the emission from the more 
optically thin CO isotopologue transitions (\S 4.3).

\subsection{Stage 1: Template Disk Structure Model}

Our starting point is with a set of disk structures calculated following the 
prescription for a steady viscous accretion disk model 
\citep[see][]{dalessio_c98,dalessio_c99,dalessio_c01,dalessio_c06}.  In these 
models, the gas surface density is determined based on the conservation of 
angular momentum flux and depends on the mass flow rate ($\dot{M}$), viscosity 
coefficient $\alpha$ \citep[]{shakura_s73}, and midplane temperature ($T_0$), 
such that $\Sigma \propto \dot{M}/\alpha T_0$.  We assume that $\dot{M}$ is 
constant throughout the disk, but make an important structural modification 
from previous generations of these models.  Motivated by the similarity 
solution for the time evolution of accretion disks \citep{hartmann_c98} and 
recent millimeter-wave observations \citep{hughes_w08}, we allow the viscosity 
coefficient to vary radially such that $\alpha = \alpha_0 \exp{R/R_c}$.  This 
change makes no difference at small radii ($R \ll R_c$), but effectively adds 
an exponential taper to the surface density profile outside the characteristic 
radius, $R_c$.  Although this modification mimics the $\Sigma$ profile derived 
from the similarity solution models, it does not reproduce the predicted 
behavior of the mass flow rate as a function of radius.  However, that 
$\dot{M}$ profile is not expected to have any observable effects on the disk 
structure at these large radii, since stellar irradiation is a much more 
important heating mechanism than viscous dissipation in those regions.  

For a given flow rate ($\dot{M}$), viscosity coefficient ($\alpha_0$), and 
characteristic radius ($R_c$), the density and temperature structure of this 
model is determined as described by \citet{dalessio_c98,dalessio_c99}. We 
consider heating from the mechanical work of viscous dissipation
(relevant only  
in the midplane of the inner disk), accretion shocks at the stellar surface, 
and passive stellar irradiation, and follow the radiative transfer of that 
energy with 1+1D calculations using the Eddington approximation and a set of 
mean dust opacities (gas opacities are considered negligible). The dust is 
assumed to be a mixture of segregated spheres composed of ``astronomical" 
silicates and graphite, with abundances (relative to the total gas mass) of 
$\zeta_{\rm sil} = 0.004$ and $\zeta_{\rm gra} = 0.0025$ \citep{draine_l84}: 
the ``reference" dust-to-gas mass ratio is $\zeta_{\rm ref} = 0.0065$.  At any 
given location in the disk, the grain size ($a$) distribution of these dust 
particles is assumed to be a power-law, $n(a) \propto a^{-3.5}$, between 
$a_{\rm min} = 0.005$\,$\mu$m and a specified $a_{\rm max}$.  We assume the 
disk has two grain populations, each with a different maximum size.  The 
``small" grains are distributed in the disk atmosphere and have $a_{\rm max} = 
0.25$\,$\mu$m (as in the interstellar medium), and the ``big" grains are 
concentrated toward the midplane and have $a_{\rm max} = 1$\,mm.
Utilizing the dust settling prescription of \citet{dalessio_c06}, the
``small" grain population in the upper disk layers is assumed to have
a dust-to-gas mass ratio $\epsilon \zeta_{\rm ref}$ and the ``big"
grain population near the midplane has a dust-to-gas ratio determined
so that mass is conserved at each radius.  A  
smooth vertical transition between the two grain populations is made at a 
height $z_{\rm big}$ above the midplane.    

The inner boundary of the structure model is taken to be the radius where 
silicates are sublimated (where $T \approx 1500$\,K).  Because this inner rim 
is irradiated by the photosphere of the central star 
\citep{natta_g00,dullemond_d00} and the accretion shock at the stellar surface 
\citep{muzerolle_d04} at a normal incidence angle, the material is heated to 
high temperatures and therefore extends (or ``puffs" up) to a larger vertical 
height, $z_{\rm wall}$.  The radial structure of that ``wall" feature (and its 
emergent spectrum) is calculated as described by \citet{dalessio_h05}.  The 
structure model is truncated at 540 AU, large enough to have no effect on our 
analysis. 

\begin{figure}[htbp]
\includegraphics[width=6in]{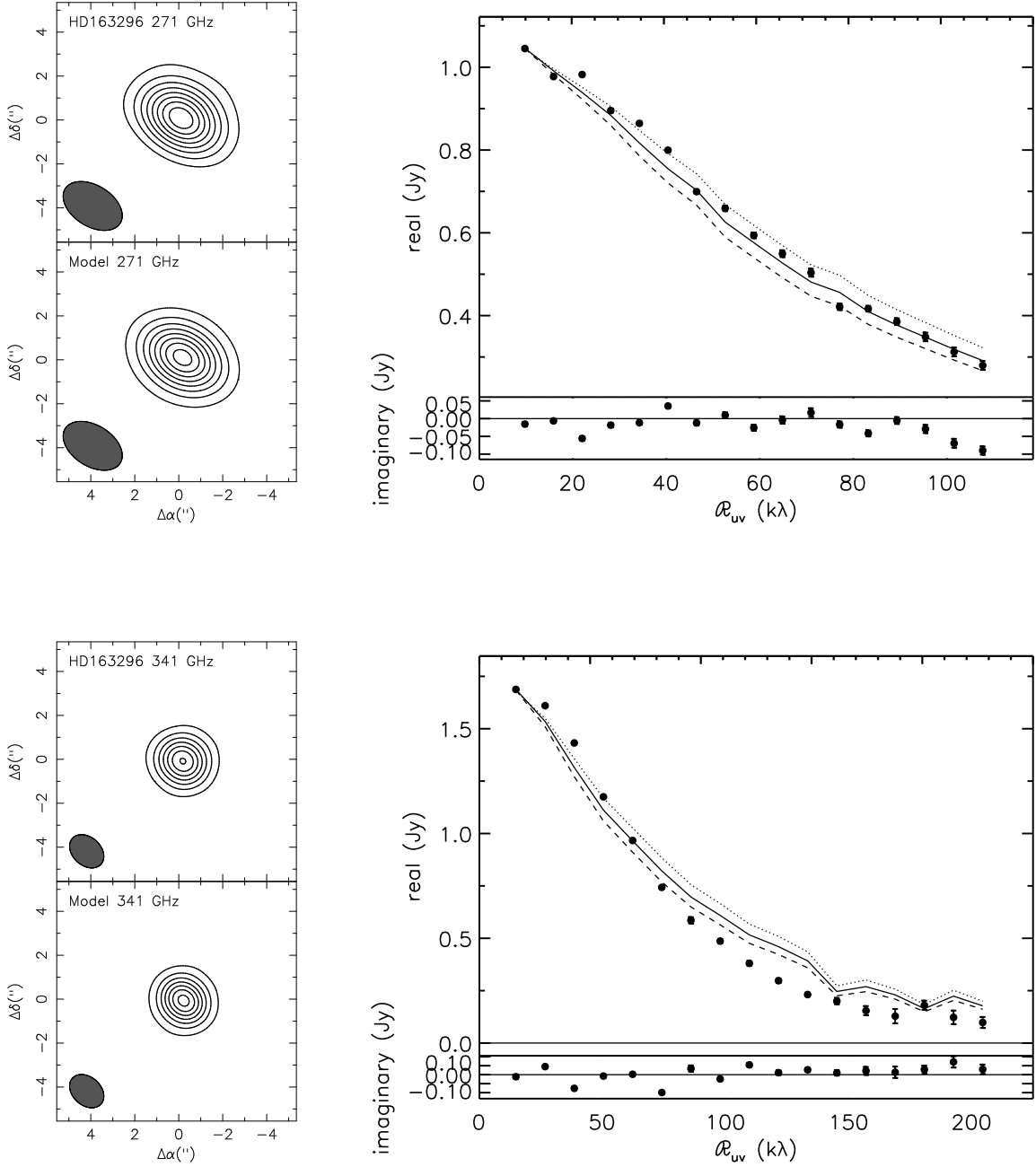}
\caption{ {\it Top panels} (Left): The predicted continuum image vs
  data at 271 GHz. Contours are shown at 3$\sigma$ intervals. (Right):
  The predicted 271 GHz continuum visibility profiles for Models with
  $R_c$=125 AU(dotted line), 150 AU(solid line), 175 AU(dashed line). 
  {\it Bottom panels:} The predicted continuum image and visibilities
  vs data at 341 GHz. 
}
\label{fig:cont}
\end{figure}

In the model calculations, we adopt the HD~163296 stellar parameters advocated
by \citet{vandenancker_d98} -- $M_{\ast} = 2.3$\,M$_{\odot}$, $R_{\ast} = 
2$\,R$_{\odot}$, and $T_{\ast} = 9333$\,K -- and a flow rate equivalent to the 
accretion rate onto the star, $\dot{M} = 7.6\times10^{-8}$\,M$_{\odot}$ 
yr$^{-1}$ \citep{garcialopez_n06}.  Based on the CO observations of 
\citet{isella_t07}, we fixed the disk inclination angle to $i = 44\degr$ and 
the major axis position angle to 133\degr\ east of north.  For this set of 
fixed parameters, the inner wall is located at a radius of 0.6\,AU.  There are 
five remaining free parameters in the model: \{$\alpha_0$, $R_c$, $\epsilon$, 
$z_{\rm wall}$, $z_{\rm big}$\}.  To identify a model structure that
provides a  
reasonable match to the HD~163296 observations, we started by comparing 
synthetic data for a coarse grid of these parameters with the broadband SED.  
That parameter search was then refined by comparing the model predictions with 
the observed millimeter continuum visibilities at 271 and 341\,GHz.  Although 
not optimized for robust parameter estimation, this approach yielded a
template  
model that exhibits a satisfactory match with these diagnostics of the dust 
disk.  The adopted model has $\alpha_0 = 0.019$, $R_c = 150$\,AU, $\epsilon = 
0.003$ (the dust-to-gas ratio in the atmosphere is 0.3\%\ of the reference 
value, $\zeta_{\rm ref}$), and $z_{\rm wall} = 0.1$\,AU.  
The total mass of the model is 0.089\,M$_{\odot}$. All the stellar and
disk properties are summarized in Table~\ref{tab:model}. 
Figure~\ref{fig:cont} shows how the
value of the $R_c$ parameter affects the corresponding continuum
visibility profiles at 271 GHz and 341 GHz, the frequencies with the
best resolved millimeter continuum data. We find the visibility
profiles are matched best with $R_c$ at 150 AU, a bit larger than
125~AU derived by \citet{hughes_w08} with the similarity solutions.  

\begin{figure}[htbp]
\includegraphics[width=5.5in]{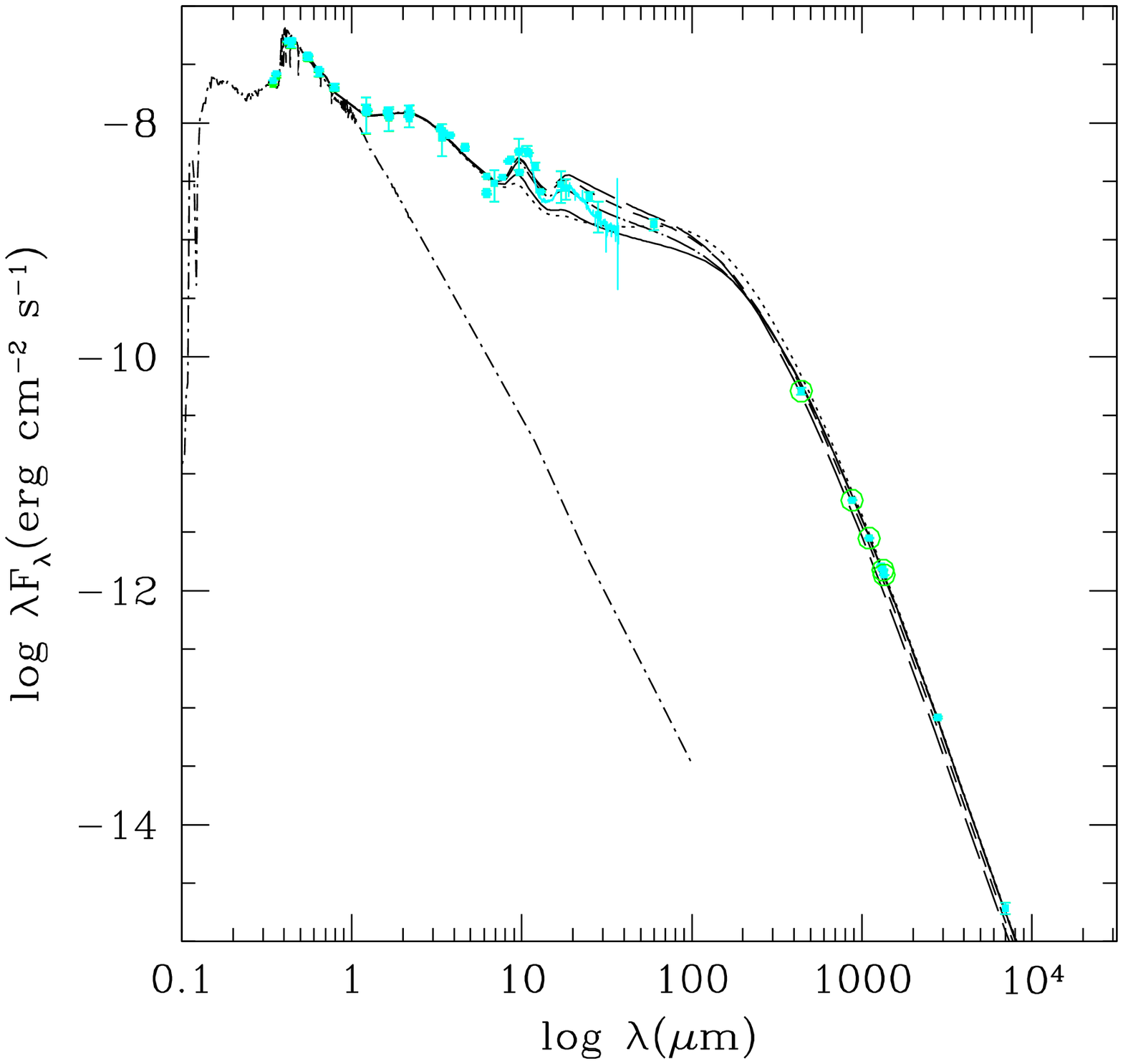}
\caption{The HD~163296 SED. Optical and infrared bands:
  UBVJHKLM from \citet{malfait_b98}; UBVRI from \citet{hillenbrand_s92};
  BV from \citet{oudmaijer_p01}; BVRIJHK from \citet{tannirkulam_m08};
  JHKLM and 8--13 $\micron$m bands from \citet{berrilli_c92}; JHK from
  \citet{eiroa_g01} and 2MASS (\citealp{cutri_s03}); ISO
  spectra (\citealp{acke_v04,thi_v01}); the CTIO and
  Keck telescopes data (\citealp{jayawardhana_f01}); IRAS
  (\citealp{beichman_n88}); Spitzer-IRS (retrieved from Spitzer archive
  website: irsa.ipac.caltech.edu). Millimeter fluxes 
  are from this paper (marked with circles) and
  \citet{isella_t07}. Overlaid on the SED are the models which
  have: $\dot{M}=7.6  \times 10^{-8} \ \MSUNYR$, $\alpha_0=0.019$,
  $\epsilon=0.003$, $R_c=150$ AU, $cos
  i=0.72$, a vertical wall at the dust sublimation radius,   
  calculated assuming $T_{sub}=1500$ K, with $R_{wall}= 0.6$ AU and a
  height $z_{wall}=0.17 R_{wall}$. The stellar photosphere (the
  short-long-dashed line) is from the library for population synthesis from
  \citet{bruzual_c93}. The model SED lines are laid out for models with
  different values of $z_{\rm big}/H=$  0.5(long-dashed),
  1.0(short-dashed), 1.5(dot-dashed), 2.0(solid) and 2.5 (dotted).  
}
\label{fig_sed}
\end{figure}

Figure~\ref{fig_sed} compares the HD~163296 SED with the models with
different values of $z_{\rm big}$, ranging from 0.5$H$ to 2.5$H$,
where $H$ is the pressure scale height calculated from the
local temperature value ($H = c_s/\Omega$, where $c_s$ is the sound
speed and $\Omega$ is the Keplerian angular velocity of the gas.  Note
that this is merely a convenient scaling, as the vertical structure is
solved consistently and does not assume a vertically isothermal profile).
The different behaviors of the models are likely
caused by the change in the shape of the irradiation surface, which
determines the fraction of stellar radiative flux intercepted and
reprocessed by the disk. Apparently it is difficult to constrain
$z_{\rm big}$ from the SED alone due to the complexity of the SED
modeling in the mid and far-IR wavelengths.
The most substantial discrepancy between the observed SED and the model 
predictions can be noted in the strength of the 10 and 18 $\mu$m silicate 
emission bands. This might be a reflection that the upper atmosphere
layers are hotter than in the  
model, or that the optical properties of the grains in the atmosphere are 
slightly different (in size distribution and/or composition) than we assume.  
The modeling formalism we are using assumes the gas and dust are in
thermal equilibrium  at all locations in the disk, even in the upper
layers of the atmosphere.   
While that is expected to be a good approximation at the height where most of 
the stellar radiation is deposited, and where most of the emission
from the bands is produced, the gas at larger heights may be even hotter
\citep[e.g.,][]{glassgold_n04, kamp_d04, jonkheid_f04,
  nomura_a07,gorti_h08, woitke_k09, kamp_t10}, which might change the dust
density distribution, and hence the silicate emission.   

\subsection{Stage 2: Refining the Vertical Structure}

\begin{figure}[htbp]
\includegraphics[width=6in]{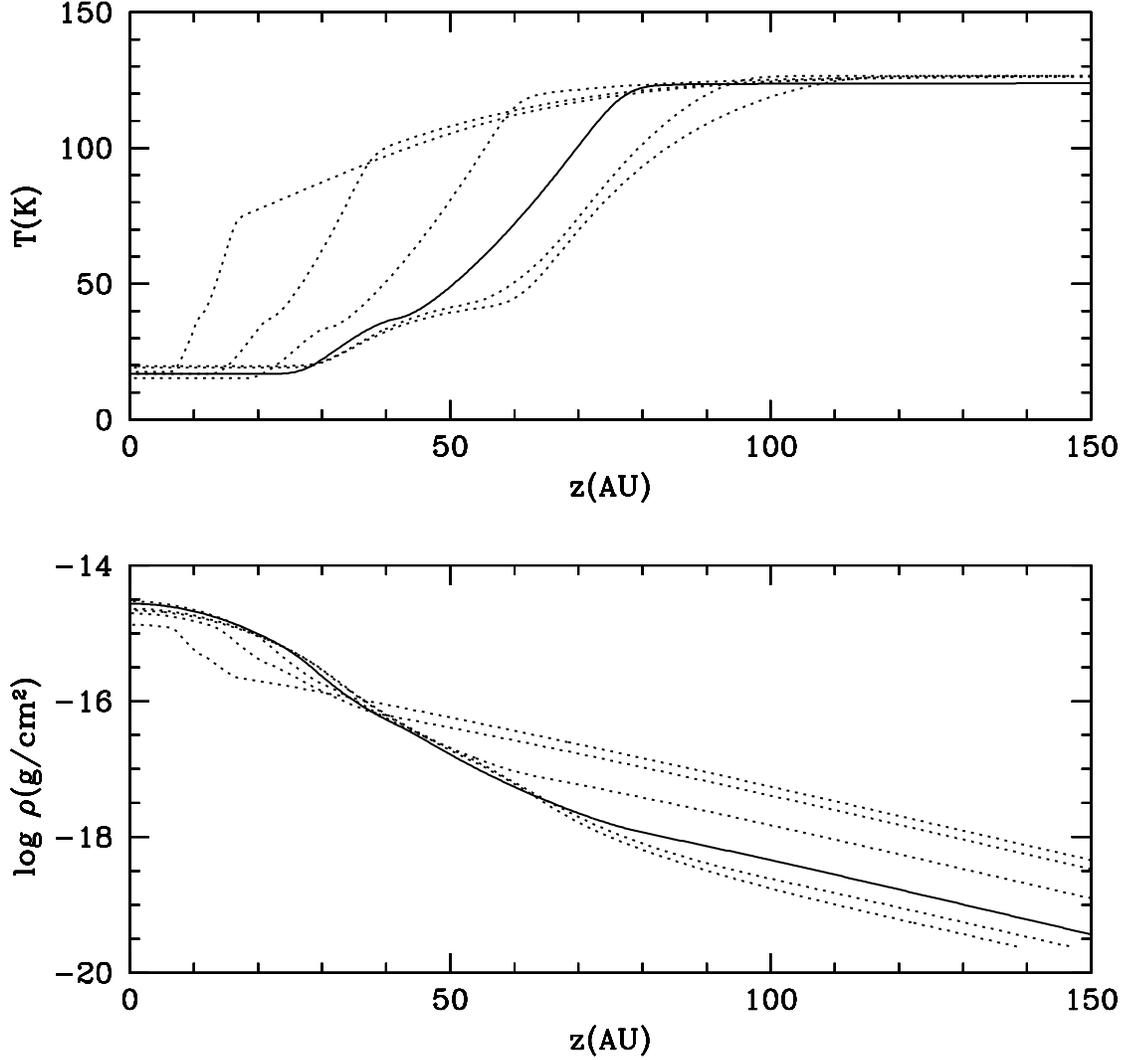}
\caption{Upper panel: Temperature vs height at $R=200$ AU for 
disk models with 
 $\dot{M}=7.6 \times 10^{-8} \ \MSUNYR$, $\alpha_0=0.019$, 
$\epsilon=0.003$, $R_c=150$ AU, and different 
values of $z_{\rm big}/H=$  0.5, 1, 1.5, 2.0, 2.5, 3 (from left to right). The fiducial model,
with $z_{\rm big}=2 H$ is showed with a solid line.
Lower panel: Density vs height at $R=200$ AU for the same models. The
lines are laid out from top to bottom for models with different 
values of $z_{\rm big}/H=$  0.5, 1, 1.5, 2.0, 2.5, 3 at $z >$ 100 AU.
}
\label{fig_temp_dens}
\end{figure}

At the large disk radii probed by our CO observations, the vertical structure 
of the disk is determined by irradiation.  The small dust grains in the disk 
atmosphere absorb energy from the incident stellar radiation field,
and some of that energy is then re-emitted down into the disk interior
at longer, infrared wavelengths.  Given the increasing densities near
the disk midplane, this ``external" heating of the disk surface
naturally produces a structure with a  
vertical temperature inversion: the midplane is cooler than the atmosphere 
\citep{calvet_p91,cg97,dalessio_c98}.  In our model prescription for dust 
settling, the deeper layers in the disk are populated by big dust grains that 
have low infrared opacities, and therefore are heated less efficiently than 
their smaller counterparts.  Therefore, this concentration of big grains near 
the disk midplane actually amplifies the temperature contrast between the 
surface and interior.  In practice, the location of the transition between the 
small and big grains, $z_{\rm big}$, has a pronounced effect on the vertical 
temperature gradient (and therefore the vertical density structure).
In Figure~\ref{fig_temp_dens}, we demonstrate how $z_{\rm big}$
impacts the vertical distributions of temperatures and densities at a
fixed radius of 200 AU.  A more condensed population of big grains
(lower $z_{\rm big}$) permits more heating  
at deeper depths into the disk (i.e., lower $z$), producing a warmer disk 
interior.  Likewise, a more vertically extended population of big grains 
(higher $z_{\rm big}$) produces much lower temperatures in the disk
interior.   

The effects of $z_{\rm big}$ are difficult to distinguish from the
dust tracers alone: the infrared SED and millimeter visibilities do
not effectively probe the shape of the vertical temperature profile. 
However, the CO line emission  
is expected to be generated in an intermediate disk layer that should be 
sensitive to this temperature inversion.  The main isotope CO lines are 
optically thick, and therefore excellent temperature diagnostics.  By
measuring the emission from several CO rotational transitions, their
relative strengths may be used to trace the temperature at different
depths in the disk (each line  
probing a layer commensurate with its excitation).  Previous analysis of 
multi-transition CO data have been used to successfully measure the
vertical temperature gradient \citep[e.g.,][]{dartois_d03}.  In this
stage of the modeling, we utilized the resolved CO 2--1, 3--2, and
6--5 emission lines toward the HD~163296 disk to determine the value
of $z_{\rm big}$ that best reproduces the inferred temperature
gradient present in its intermediate layer. 

To compare the model predictions with the CO data, we first need to
fix a set of geometric and kinematic parameters that affect the observed
spatio-kinematic  
behavior of the disk.  We assume the disk material orbits the central star in 
Keplerian motion, and fix the stellar mass and position, and non-thermal 
turbulent velocity width ($dV_{turb}$) based on the models of 
\citet{hughes_w11} and the references noted in \S 4.1. The effects of
inclination ($i$), position angle of the major axis (PA), and systemic
velocity ($V_{lsr}$) are essentially orthogonal 
to those that depend on the assumed disk structure, and therefore can be 
optimized independently. All the disk geometric and kinematic
parameters are summarized in Table~\ref{tab:model}. We further assume
a simple CO abundance model that  
depends on the local density and temperature via the parameters introduced by 
\citet{qi_w08}, described below. For a given disk structure and CO
abundance model, we use the 
non-local thermodynamic equilibrium two-dimensional accelerated
Monte-Carlo radiative transfer 
code {\tt RATRAN} to calculate the molecular excitation and generate a
sky-projected set of synthetic CO data cubes \citep{hogerheijde_v00},
sampled at the same spatial frequencies as each SMA dataset. The
collisional rates are taken from the Leiden Atomic and Molecular
Database \citep{schoier_v05} for non-LTE line radiative transfer
calculations. Specifically, in the models used in this paper, we have
used the new set of the CO collisional rate coefficients calculated by
\citet{yang_s10}.  

Our adopted abundance model was introduced by \citet{qi_w08}, and 
assumes that the CO emission originates in a vertical layer of the disk with a 
constant abundance.  The upper (surface) and lower (midplane) boundaries of 
that layer are defined by the parameters $\sigma_s$ and $\sigma_m$, which 
represent vertically-integrated hydrogen column
densities from the disk surface in units of
$1.59\times10^{21}$\,cm$^{-2}$ (the conversion 
factor of the hydrogen column to $A_v$ for interstellar dust).  
The CO abundance ($f_{\rm CO}$) is assumed to be a constant
for the hydrogen column densities from the disk surface $\sigma_s \ge
N \ge \sigma_m$, with respect to the H$_2$ density  
specified in the disk structure model. 

\begin{figure}[htbp]
\includegraphics[width=6in]{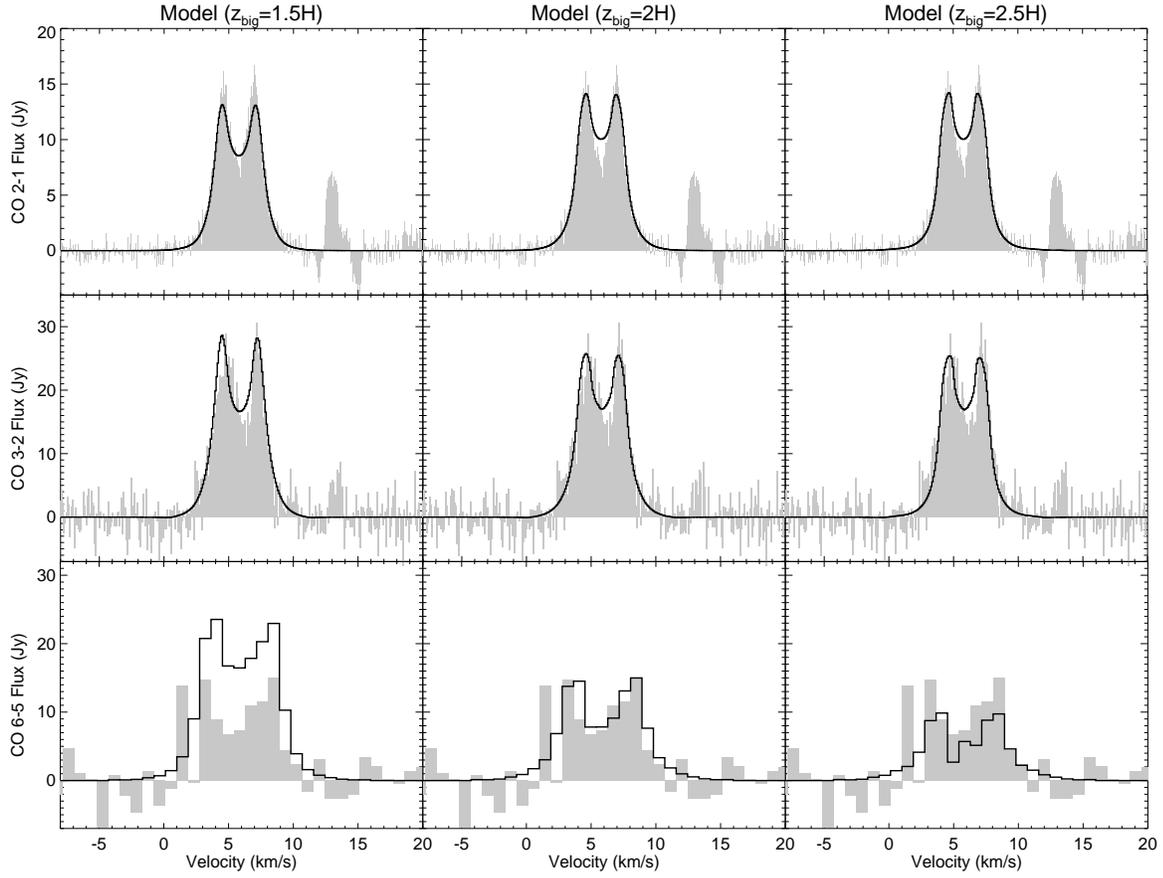}
\caption{ CO J=2--1, 3--2, and 6--5 model spectra (black lines)
  overlaid with the HD 163296 spectra in grey shade. The left, middle
  and right column are the simulated models with $z_{\rm big}/H=$ 1.5,
  2.0, 2.5. Note that all the models have been sampled at the same
  spatial frequencies as each SMA dataset. 
}
\label{fig:modelspec}
\end{figure}

\begin{figure}[htbp]
\includegraphics[width=2.2in,angle=90]{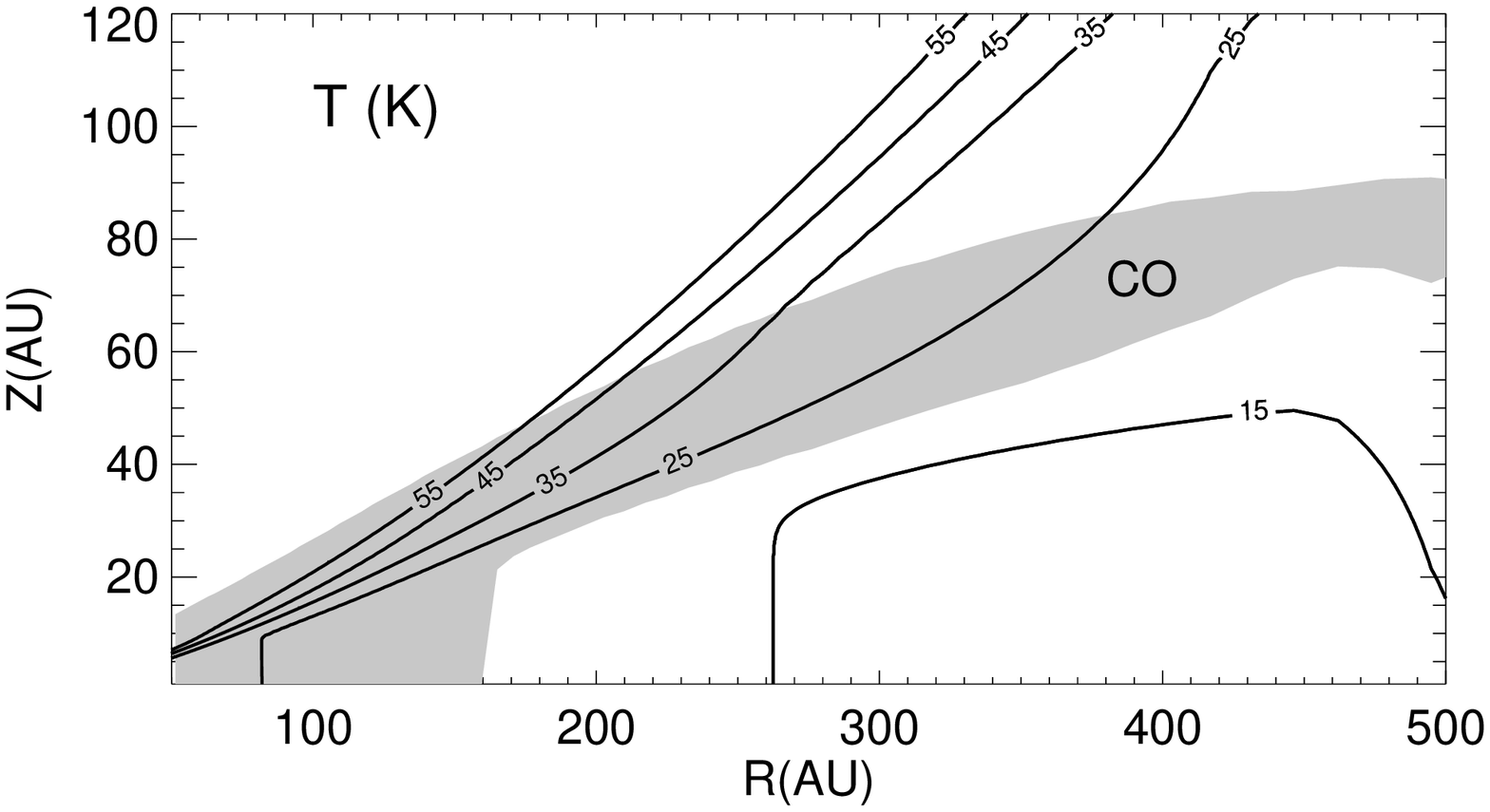}
\vskip 5mm
\includegraphics[width=2.2in,angle=90]{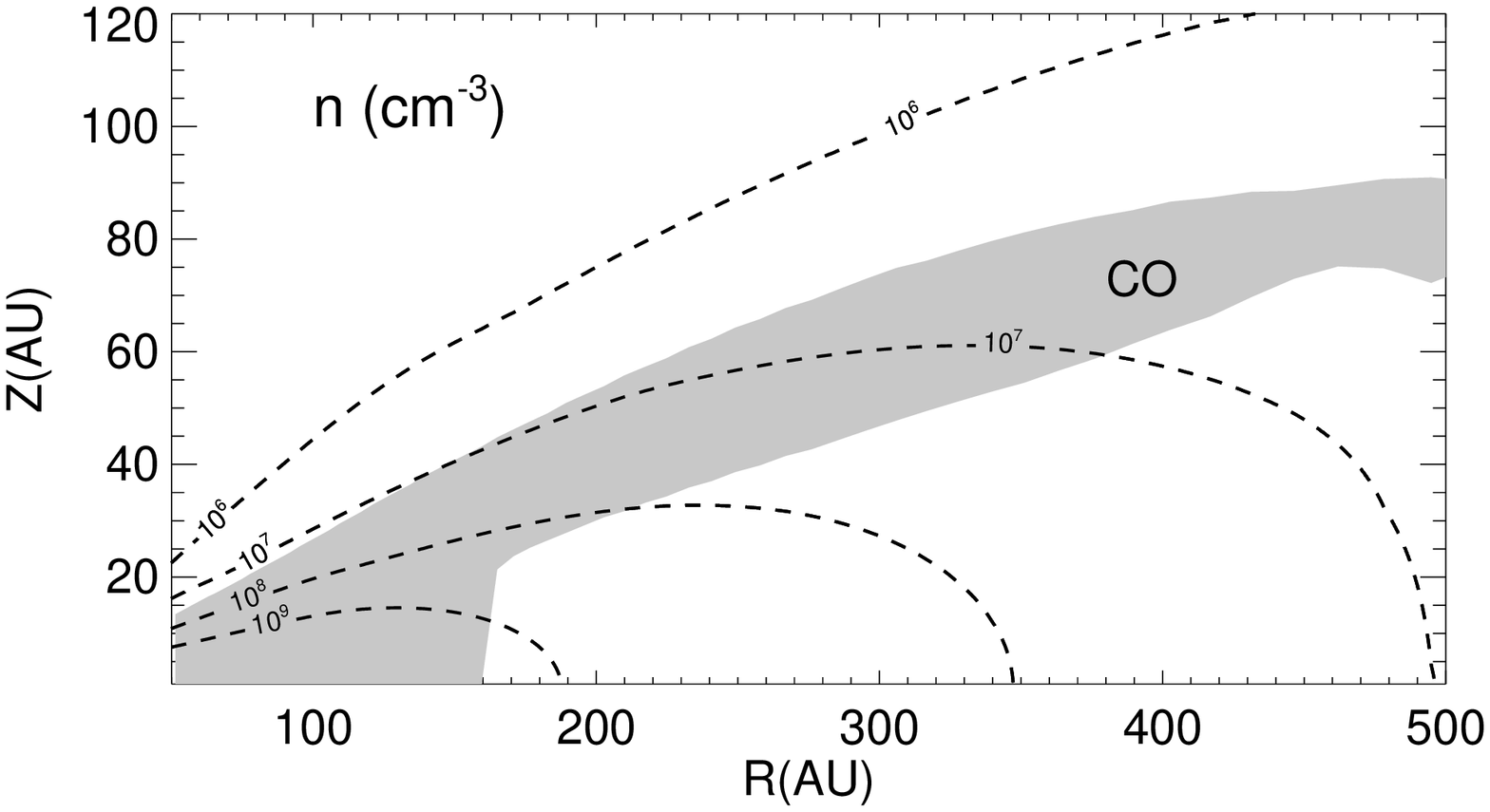}
\vskip 5mm
\includegraphics[width=2.2in,angle=90]{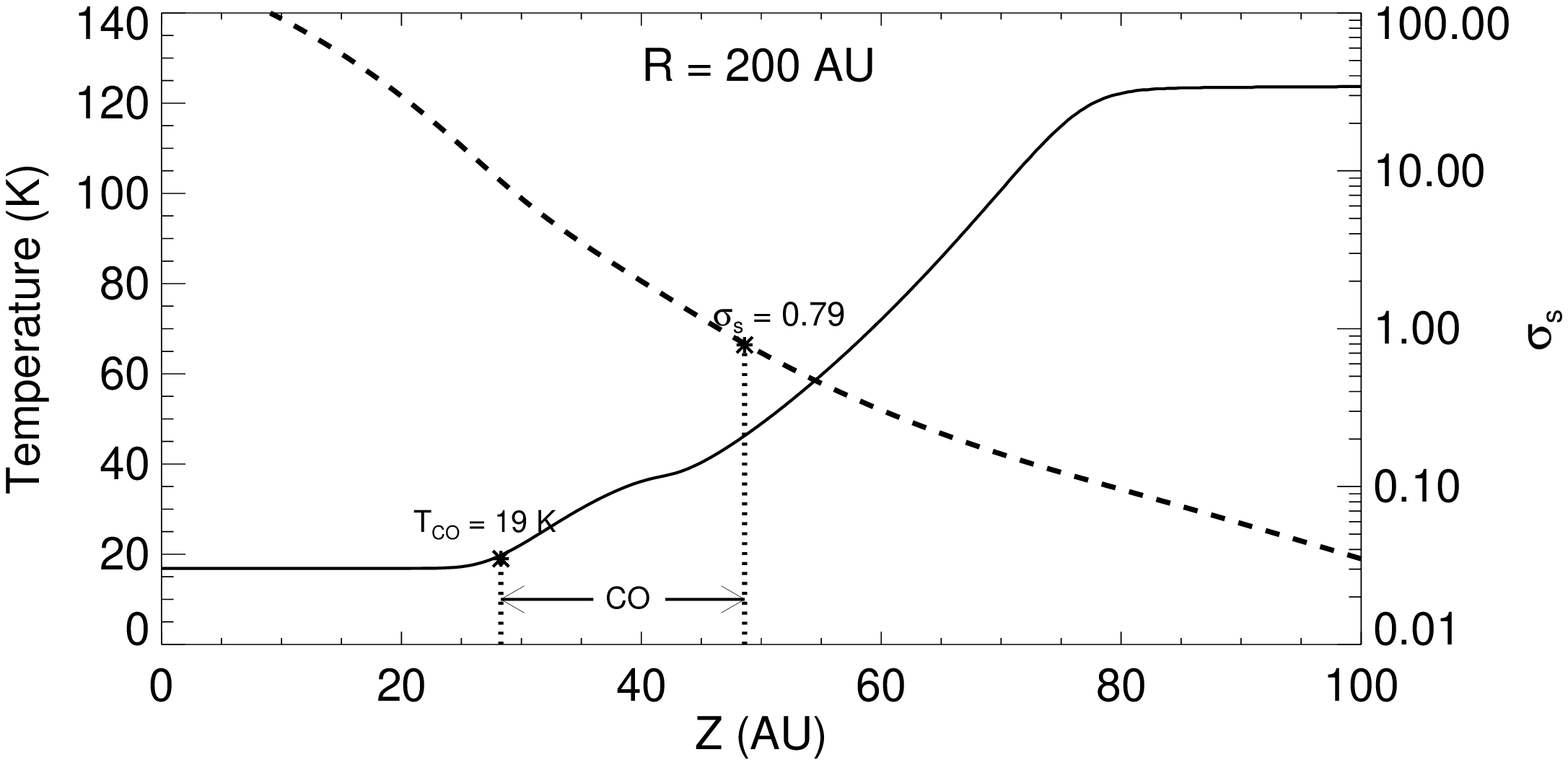}
\caption{ The irradiated accretion disk structure model for HD~163296
 with $z_{\rm big}/H=$ 2.0 and $R_c$=150 AU. {\it Top and middle
 panels}: Temperature and density profiles  are indicated by solid and dashed
  lines, respectively. The CO emission area constrained by the best-fit
 vertical boundaries is shown in grey shade. {\it
 Bottom panel}: The vertical distributions of temperature and
 $\sigma_s$ at R = 200 AU are indicated by solid and dashed lines,
 respectively. The vertical dotted lines show the best-fit locations of the
 lower boundary (T$_{\rm CO}$ = 19 K) and upper boundary
 ($\sigma_s$=0.79) for CO.   
}
\label{fig:structure}
\end{figure}

We compute model CO visibilities for a range of disk structure models with 
different $z_{\rm big}$ values, ranging from 0.5--3.0$H$ in 0.5$H$
steps, where $H$ is the pressure scale height as we described before.
For each $z_{\rm big}$ 
value, we optimized the CO abundance model ($\sigma_m$, $\sigma_s$ and
fractional abundance $f_{\rm CO}$) to minimize the 
combined $\chi^2$ value in reference to the CO J=2--1 and 3--2 visibility 
data. Because the sensitivity of our CO J=6--5 data is relatively
modest, this  
high-excitation line is better suited to an {\it a posteriori} check on the 
fitting results.  Nevertheless, the J=6--5 line has a sufficiently high 
excitation that its luminosity relative to the low-lying J=2--1 and 3--2 lines 
provides perhaps the best discriminant between different $z_{\rm big}$
values. Figure \ref{fig:modelspec} shows the results of this stage of
the modeling analysis, marking a direct comparison between the
observed and model spectra of the main CO transitions for 3 different
$z_{\rm big}$ values - 1.5, 2.0, and 2.5$H$.  While any of these
models provides a suitable match to both the J=2--1 and 3--2 spectra,
only the $z_{\rm big} = 2H$ model also matches the J=6--5 line. At R=100
AU of this model, the abundance of big grains has decreased to 50\%\ of
its maximum value at a height around 16 AU from the midplane.  
Figure~\ref{fig:structure} shows our favored two-dimensional density
and temperature distributions for this $z_{\rm big} = 2H$ model.  

\subsection{Stage 3: CO Abundance Distributions} 

We have established a gas+dust structure model that reproduces 
well the broadband SED, resolved millimeter continuum images, and a 
multi-transition set of CO spectral images.  The next step in the modeling 
analysis is to constrain the spatial distribution of the CO abundance, relying 
on the spatially resolved - and more optically thin - CO isotopologue 
emission, which probe much deeper into the midplane. Here we assume CO
and its isotopologues share the same spatial distribution and only
differ in fractional abundances. Instead of
$\sigma_m$, at this stage we adopt the lower boundary (toward midplane)
governed by the CO 
freeze-out temperature, T$_{\rm CO}$, such that $f_{\rm CO}
\rightarrow 0$ when $T < T_{\rm CO}$, which provides a
way of interpreting this lower boundary more physically. For the
upper boundary (toward surface), we still adopt $\sigma_{s}$ as
before. 

\begin{figure}[htbp]
\includegraphics[width=2.5in,angle=90]{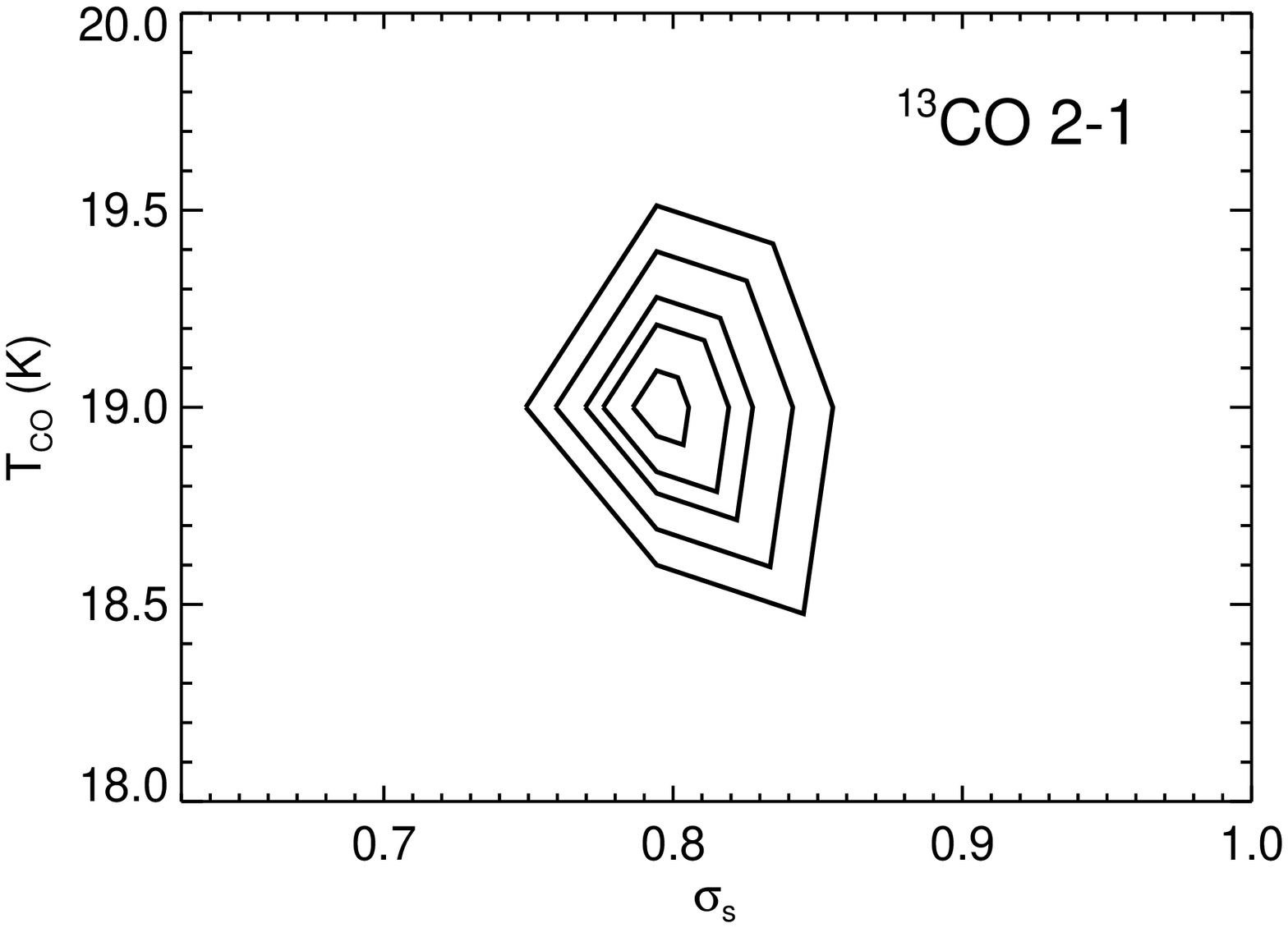}\\
\vskip 5mm
\includegraphics[width=2.5in,angle=90]{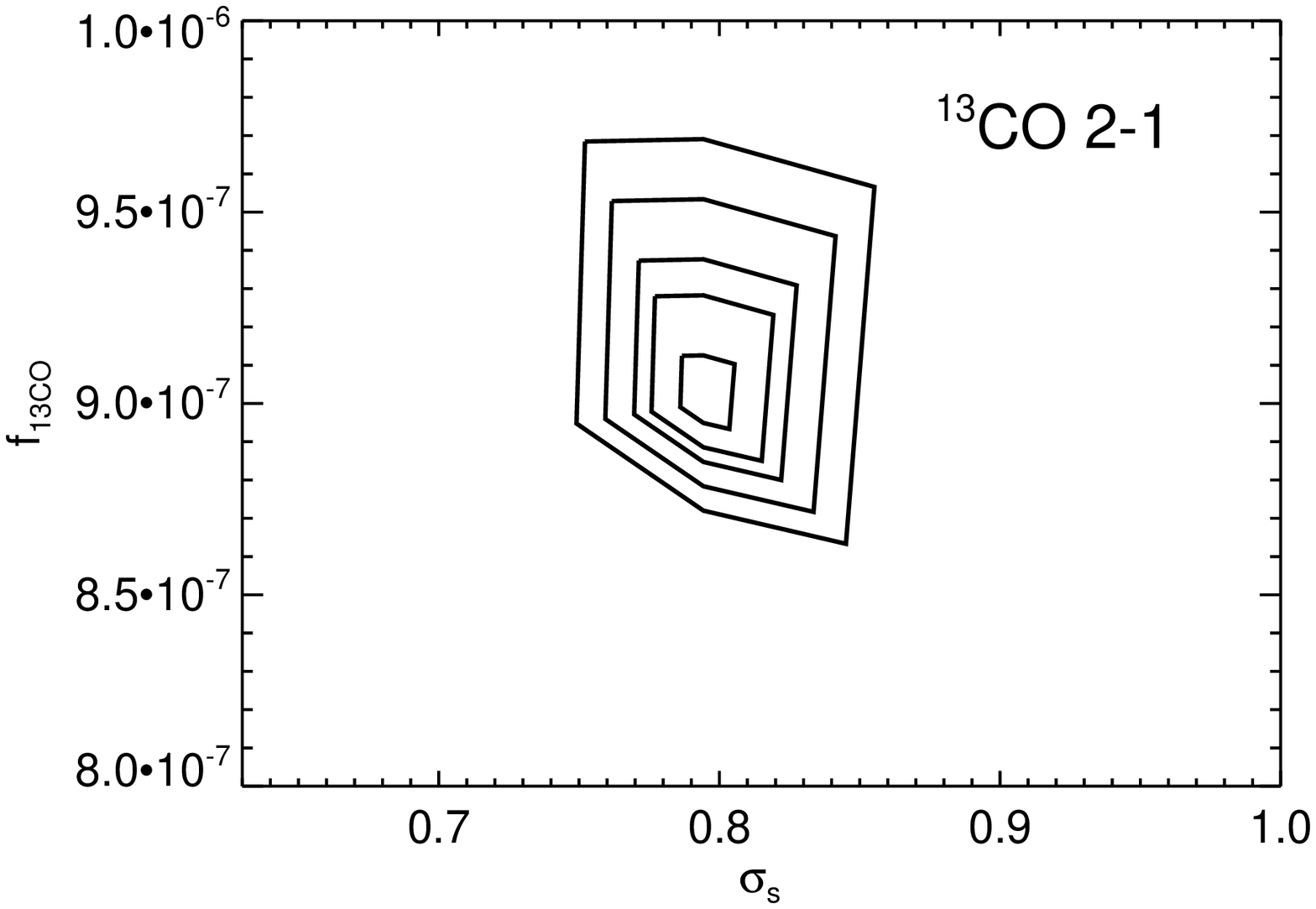}\\
\vskip 5mm
\includegraphics[width=2.5in,angle=90]{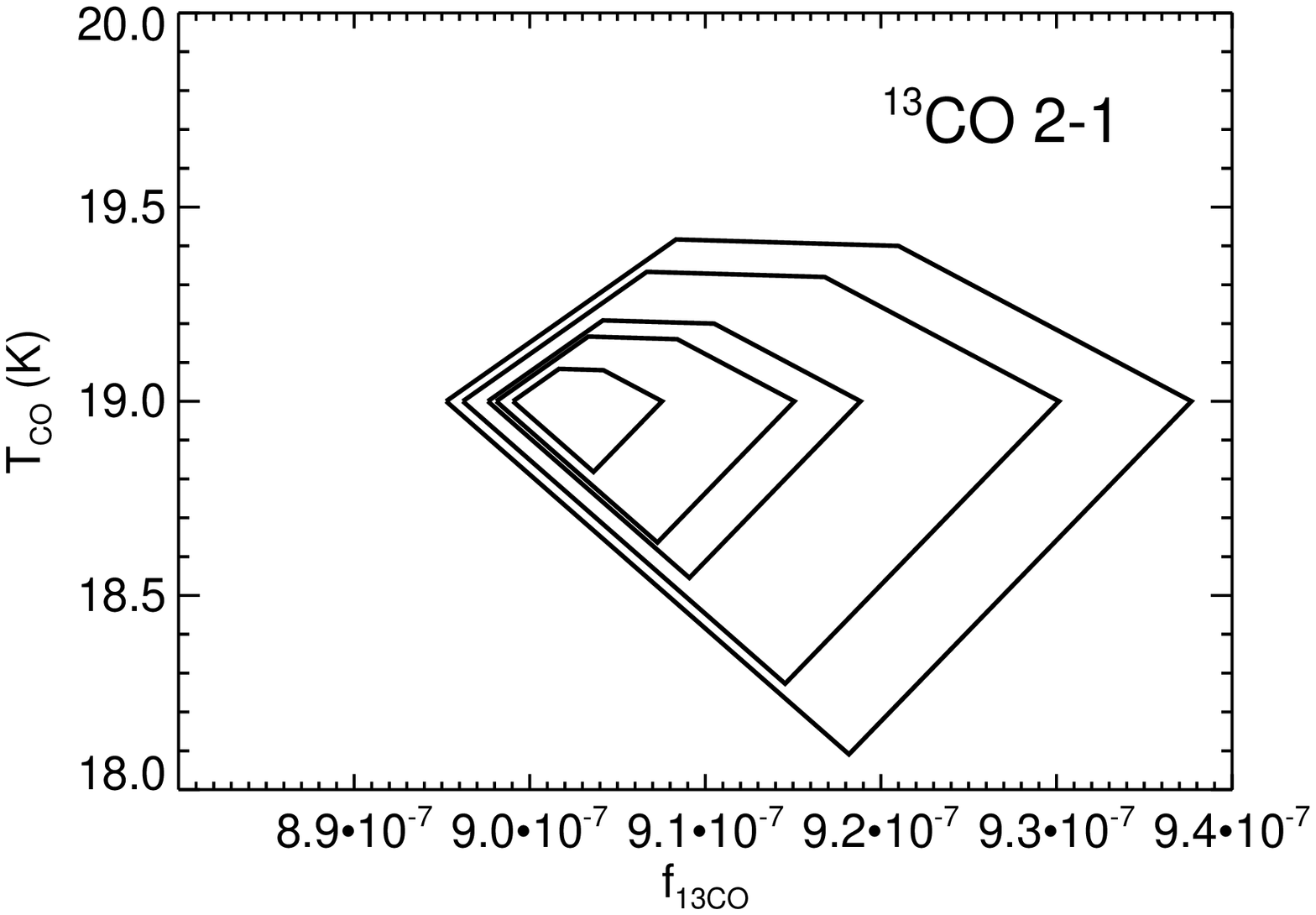}
\caption{Iso-$\chi^2$ surfaces of CO freeze-out temperature (T$_{\rm
    CO}$), CO fractional abundance (f$_{\rm CO}$) and $\sigma_s$ for
    $^{13}$CO. Contours correspond to the 1--5 $\sigma$ errors.    
}
\label{fig:zcentco}
\end{figure}

Using the fixed structure model derived in \S 4.1 and 4.2, we compute
a grid of  
synthetic $^{13}$CO J=2--1 visibility datasets over a range of $\sigma_s$, 
$T_{\rm CO}$, and $f_{\rm ^{13}CO}$ values and compare with the
observations. Figure~\ref{fig:zcentco} shows the $\chi^2$ surfaces for
the $^{13}$CO J=2--1 emission in the space of these three parameters. We 
find that the data are best described when $\sigma_s = 0.79 \pm 0.03$ (i.e., 
the hydrogen column densities from the disk surface are below
1.2--1.3$\times10^{21}$ cm$^{-2}$), $T_{\rm CO}  
= 19.0\pm0.3$ K, and $f_{\rm ^{13}CO} = 9.0(\pm0.6)\times10^{-7}$.
Figure~\ref{fig:structure} shows the locations of the CO emission in
gray shade on top of the temperature and density profiles (top and
middle panels) and the two vertical boundaries ($\sigma_s = 0.79$ and $T_{\rm
  CO} = 19 $K) at R = 200 AU (bottom panel). 

\begin{figure}[htbp]
\includegraphics[width=4in,angle=90]{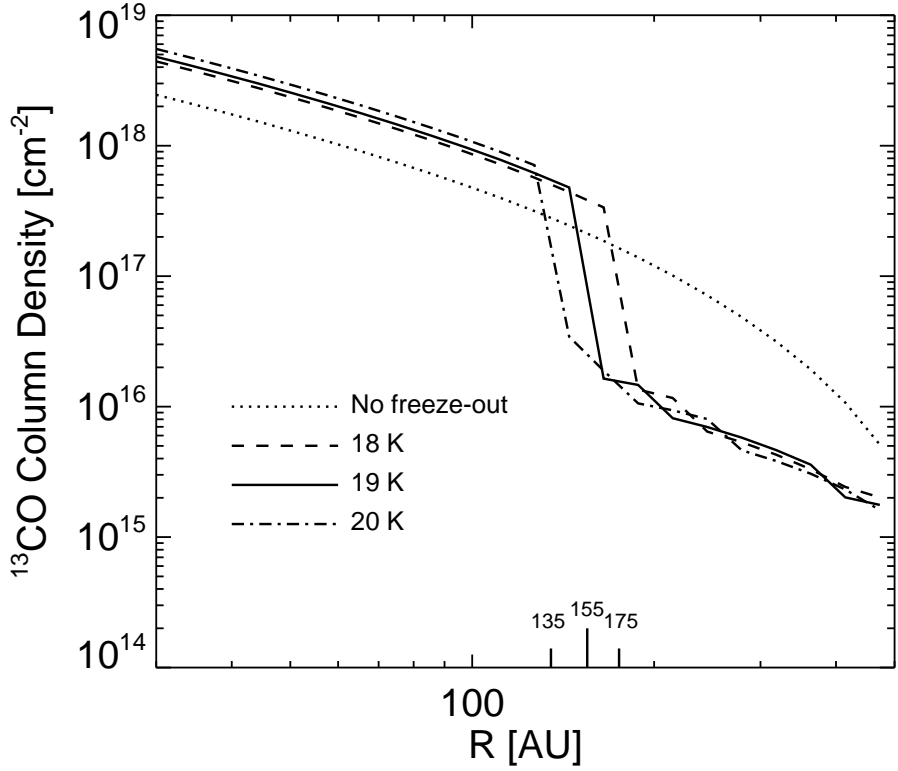}
\caption{Models of $^{13}$CO radial column densities for no freeze-out
  (dotted), freeze-out at 18~K (dashed) 19~K (solid) and 20~K
  (dot-dashed). Each model has been scaled to fit the $^{13}$CO~2--1
  emission. Note that the CO snow line is at a radius of 155 AU for
  T$_{\rm CO}=$ 19 K and it increases from 135 to 175 AU when T$_{\rm CO}$
  decreases from 20 to 18 K. 
}
\label{fig:col-13co}
\end{figure}

Figure~\ref{fig:col-13co} shows the effect of the CO freeze-out
temperatures T$_{\rm CO}$ on the radial column density of $^{13}$CO in the
model. Without any freeze-out, the $^{13}$CO column density follows
the exponential taper of the H$_2$ density profile. The freeze-out at
19 K leads to a significant drop in the gas-phase $^{13}$CO column
density beyond a radius of $\sim$155 AU (or 310 AU in diameter), which
we directly resolve. Even though the $^{13}$CO column
densities are different by orders of magnitude , the line emission
difference projected to the line-of-sight 
from the outer disk is very subtle and high spectral resolution data is
fundamental for resolving any molecular abundance structure changes at
radii beyond $\sim$155 AU. Indeed a similar analysis of the C$^{18}$O
J=2--1 emission 
alone indicates that the data can be fit equally well (or perhaps
better) by models that do not include CO freeze-out (i.e. the lower
boundary is the midplane, $z = 0$). However, this apparent
inconsistency is likely the result of the 10$\times$ lower velocity
resolution of the C$^{18}$O data. Therefore, we don't fit for CO vertical
boundaries from the emissions of C$^{18}$O 2--1 and C$^{17}$O 3--2 due
to their weaker signals and limited spectral resolutions.    

\begin{figure}[htbp]
\includegraphics[width=5.5in]{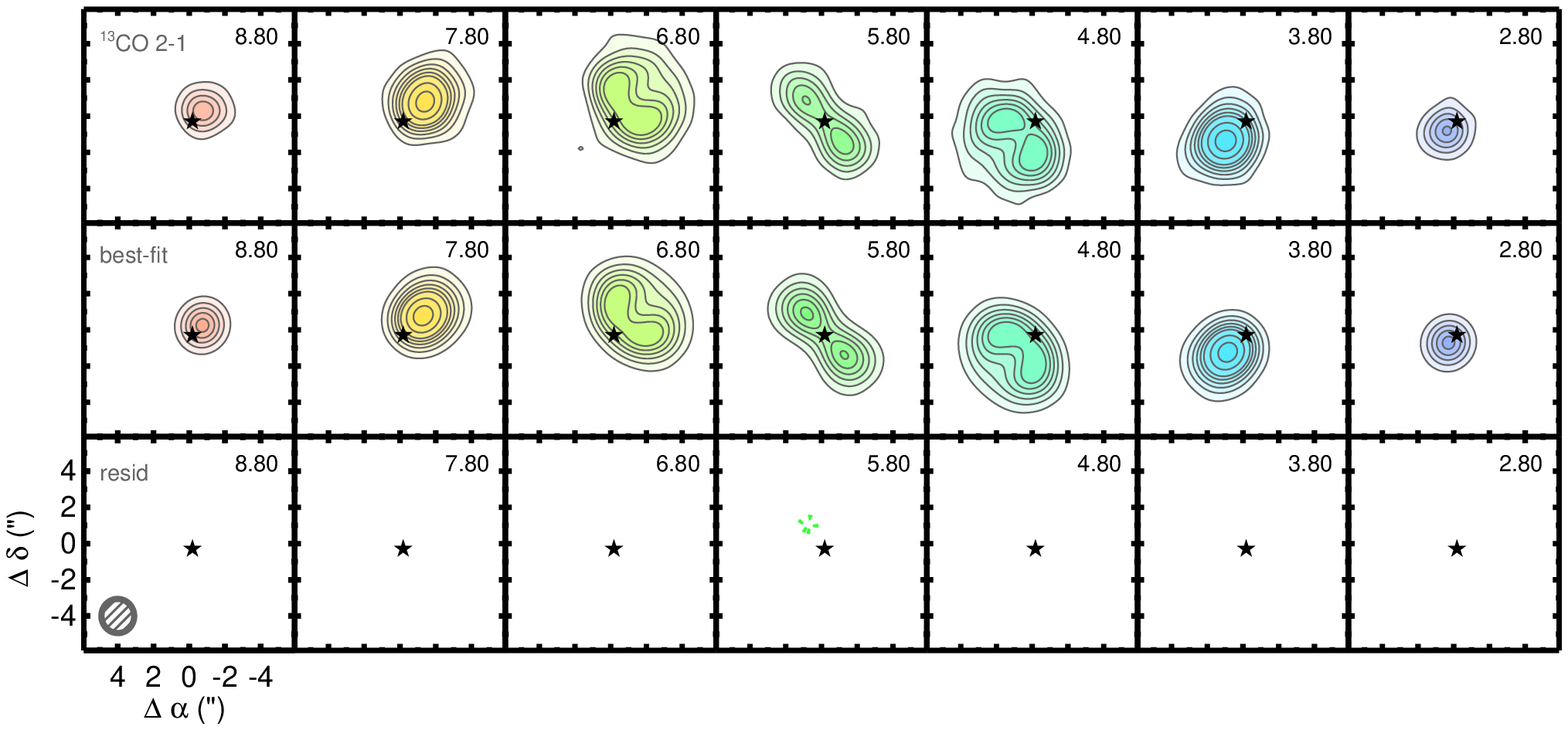}\\
\includegraphics[width=5.5in]{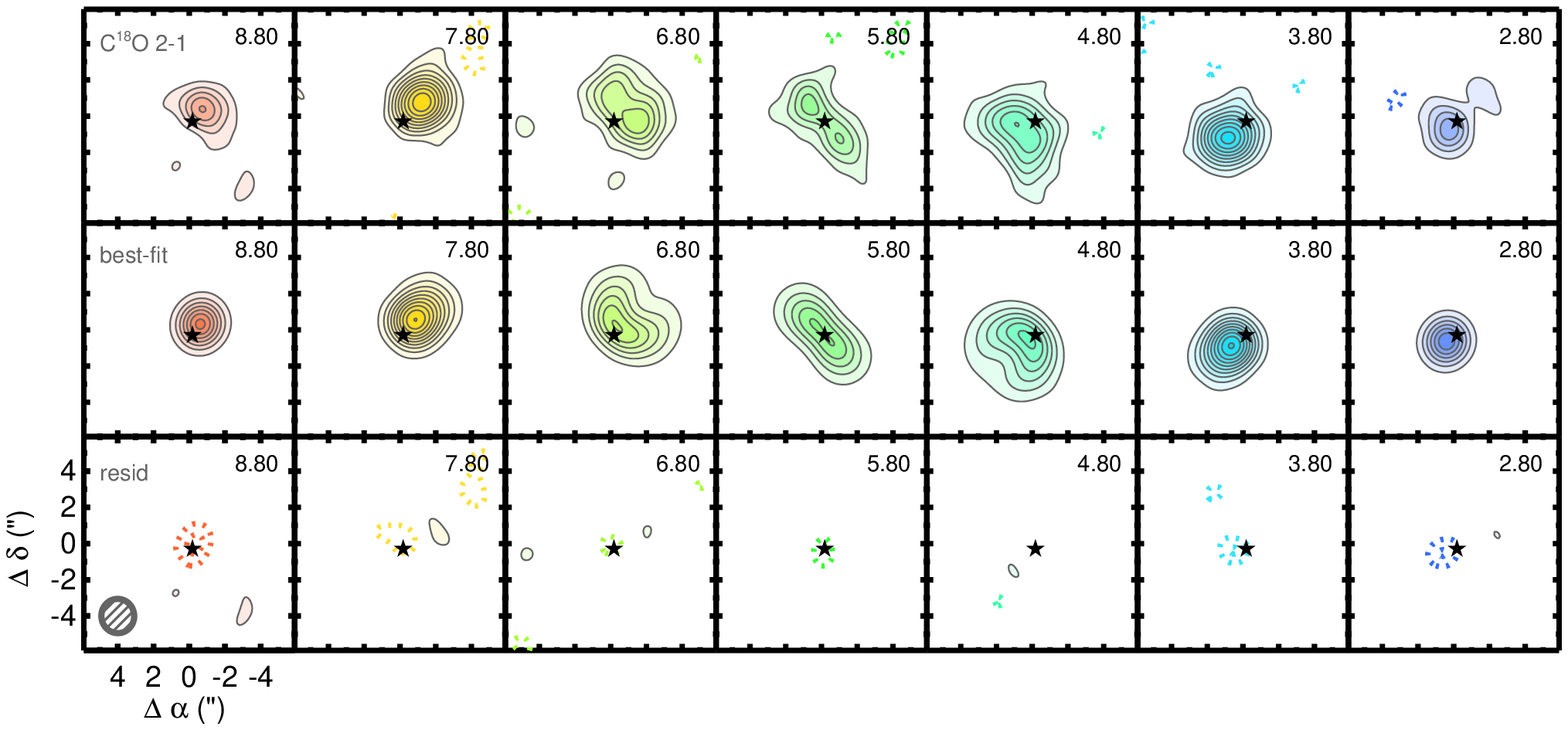}\\
\includegraphics[width=5.5in]{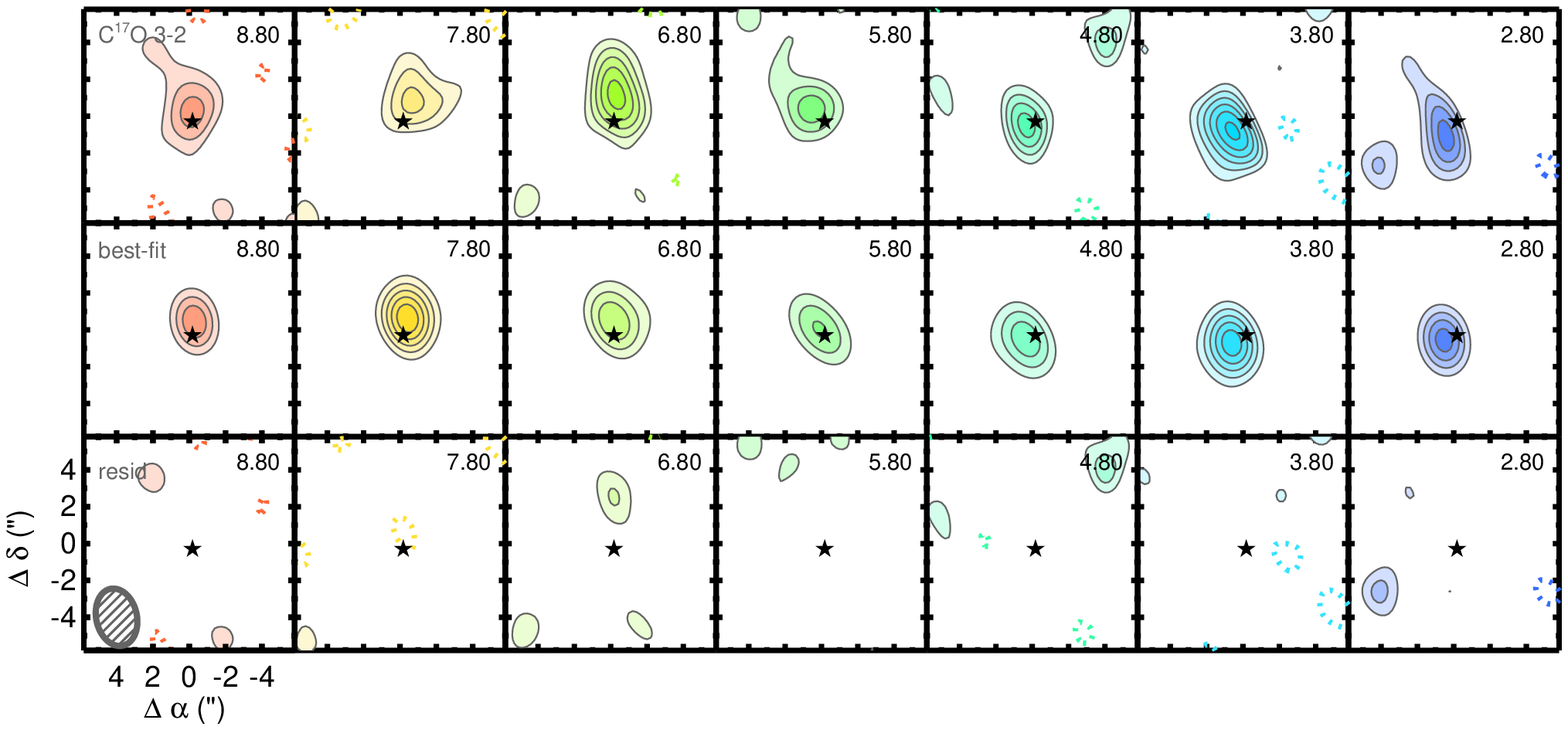}
\caption{For each panel, the top rows are the velocity channel maps of
  the $^{13}$CO, C$^{18}$O 2--1 and 
  C$^{17}$O 3--2 emissions toward HD~163296, respectively (velocities
  binned in 1 km s$^{-1}$. Contours
  are 0.04 Jy Beam$^{-1}$ (1$\sigma$) $\times
  [3,6,9,12,15,18,24,30,36,42,48,54]$ for $^{13}$CO 2--1; 0.03 Jy
  Beam$^{-1}$ (1$\sigma$) $\times [2,4,6,8,10,12,14,16,18,20]$ for C$^{18}$O
  2--1; 0.15 Jy Beam$^{-1}$ (1$\sigma$) $\times [2,3,4,5,6,7]$ for
  C$^{17}$O 3--2. The middle 
  rows are the best-fit models and the bottom rows are the difference
  between the best-fit models and data on the same contour scale. 
}
\label{fig:coisotope}
\end{figure}

\begin{figure}[htbp]
\includegraphics[width=5.5in]{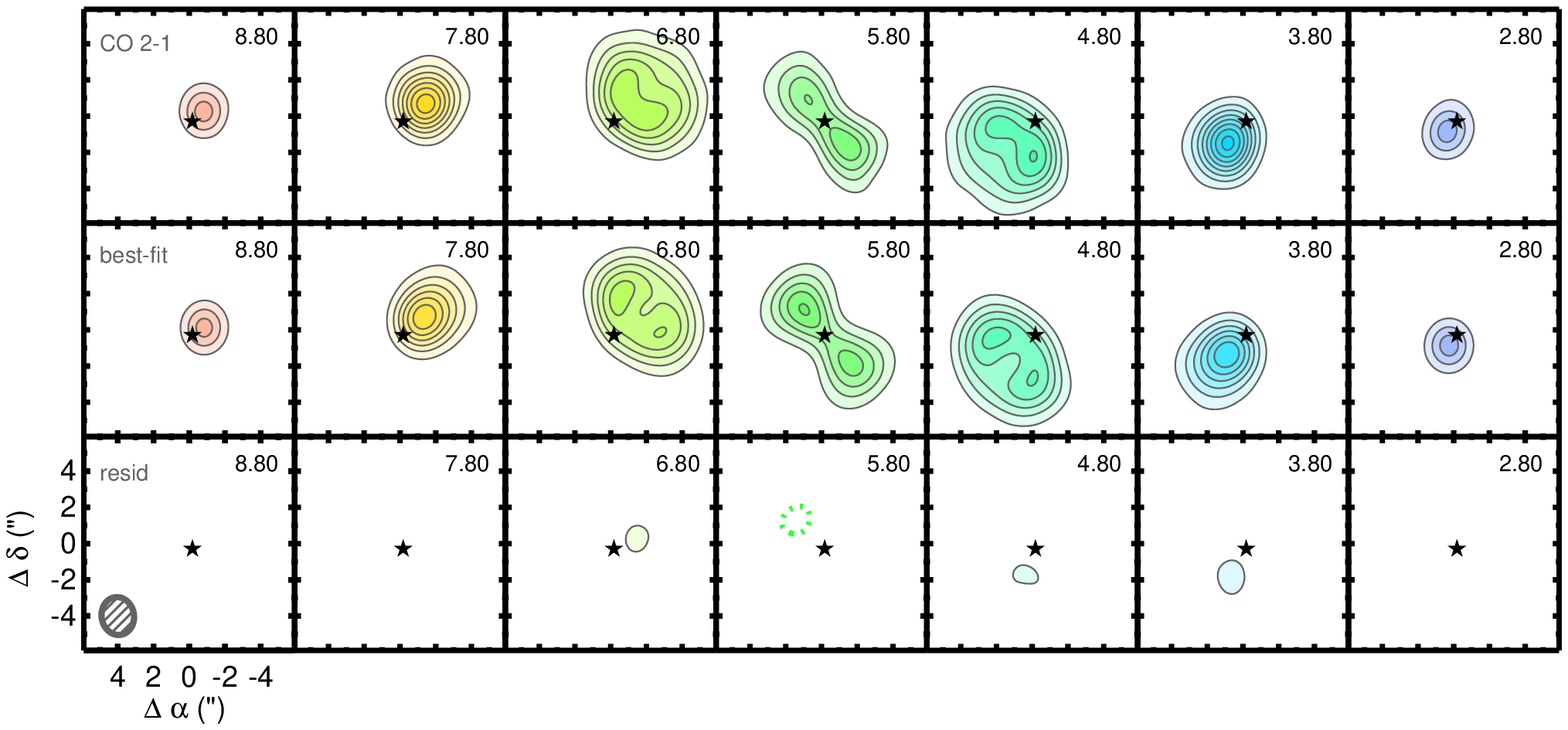} \\
\includegraphics[width=5.5in]{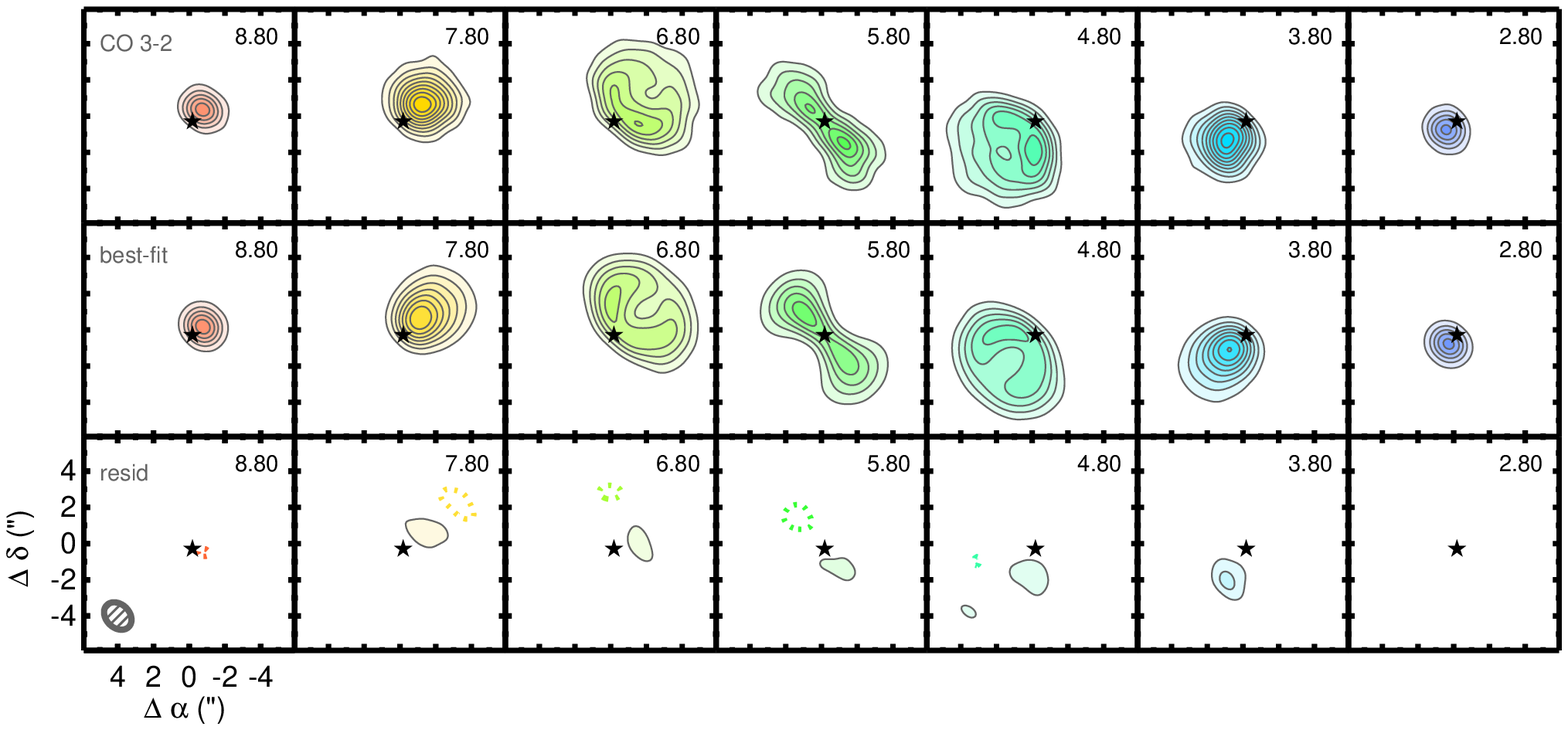} \\
\includegraphics[width=5.5in]{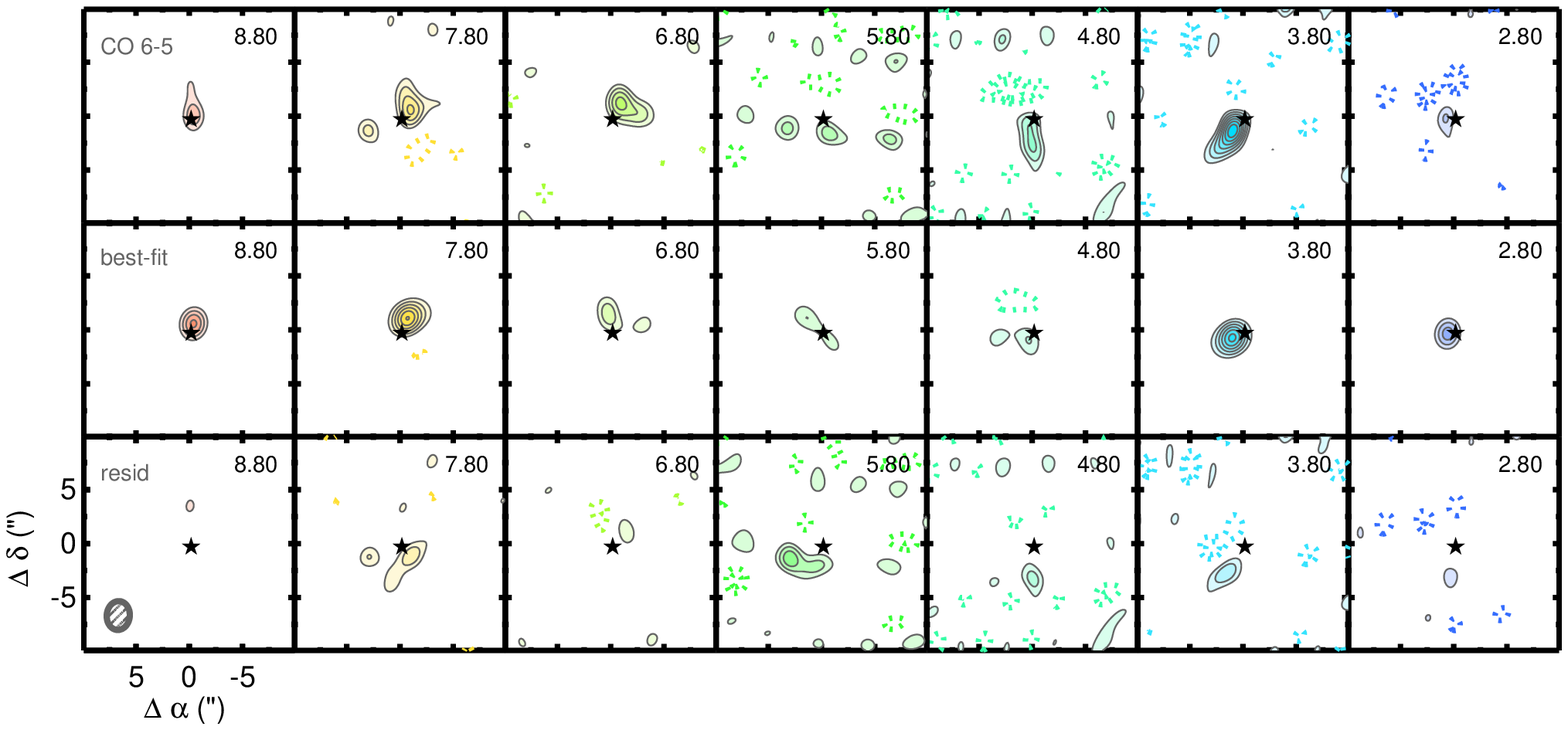}
\caption{For each panel, the top rows are the velocity channel maps of
  the CO 2--1, 3--2 and 6--5 emissions toward HD~163296,
  respectively (velocities binned in 1 km s$^{-1}$). Contours are 0.15 Jy Beam$^{-1}$ (1$\sigma$) $\times
  [3,6,9,12,15,18,21,24]$ for CO 2--1; 0.2 Jy
  Beam$^{-1}$ (1$\sigma$) $\times
  [3,6,9,12,15,18,21,24,27]$ for CO 3--2; 3.0 Jy
  Beam$^{-1}$ (1$\sigma$) $\times [2,3,4,5,6,7,8]$ for CO
  6--5. The middle  
  rows are the best-fit models and the bottom rows are the difference
  between the best-fit models and data on the same contour scale. 
}
\label{fig:cofit}
\end{figure}

Keeping that resolution effect in mind, we adopt the abundance boundaries 
derived from the $^{13}$CO 2--1 data and then fit the CO, C$^{18}$O, and 
C$^{17}$O data with only the fractional abundances as free 
parameters.  For the $\sigma_s$ and  $T_{\rm CO}$ derived above, we
find the fractional abundances of CO, C$^{18}$O and  
C$^{17}$O to be $6.0(\pm0.3)\times 10^{-5}$, $1.35(\pm0.20)\times
10^{-7}$ and $3.5(\pm1.1)\times10^{-8}$ respectively, corresponding to 
CO/$^{13}$CO = $67\pm8$, CO/C$^{18}$O = $444\pm88$ and  
CO/C$^{17}$O = $3.8\pm1.7$.  Our derived isotopic ratios are all consistent 
with the quiescent interstellar gas-phase values, which
\citet{wilson99} find in the local ISM
to be CO/$^{13}$CO = $69\pm6$, CO/C$^{18}$O = $557\pm30$, and CO/C$^{17}$O = 
$3.6\pm0.2$.  Our final model parameters are listed in
Table~\ref{tab:model2}.  The 
models are directly compared with the data channel maps in Figures
\ref{fig:coisotope} and Figure \ref{fig:cofit} (with the velocities
binned in 1 km s$^{-1}$ channels). Table~\ref{tab:model2} also shows the
best-fit fractional abundances in a model  
of HD~163296 that does not include CO freeze-out; the $^{13}$CO/C$^{18}$O 
ratio is determined to be $30.6\pm6.1$, about four times higher than the
value $6.7\pm1.4$ derived from our best-fit models that  
consider CO freeze-out at 19~K (Table~\ref{tab:model2}) or the local
ISM value $8.1\pm1.1$ (\citealp{wilson99}). This provides an
indirect evidence that we should also take into account CO freeze-out
in the C$^{18}$O 2--1 data analysis. 

\section{Discussion}

\subsection{Modeling Dust Emission and CO Line Emission}

We have modeled multiple emission lines of CO and its isotopologues from the 
disk around HD~163296 in the context of an accretion disk model structure,  
grounded in observations of the broadband SED and resolved millimeter
continuum  
emission.  The goal of this modeling effort is to develop a more consistent 
examination of the connection between the gas and dust phases in the disk.  
While our modeling framework does not treat the complexity of a completely 
self-consistent, simultaneous description of the energy balance and chemistry 
between the gas and dust, it effectively employs a set of parameters that can 
retrieve molecular abundance information in a way that captures the essential 
character of the layered disk structures predicted by those more sophisticated 
models.  Most importantly, our approach directly addresses two common issues 
that are noted in much simpler structure models: (1) a radius (or size) 
discrepancy between the dust and CO emission, and (2) the degeneracy in the 
vertical temperature structure for models based solely on the dust emission 
(i.e., the SED).  

A longstanding problem with disk models has been the seemingly different 
radial distributions of dust and gas when each is considered
independently (\citealp{dutrey_g07}). \citet{isella_t07} presented
multi-wavelength millimeter continuum and CO isotopologue observations
of the disk around HD~163296 and found a significant discrepancy
between the outer radius derived for the dust continuum ($200 \pm 15$
AU) and that derived from CO emission ($540 \pm 40$ AU) in their
truncated power law models. However, \citet{hughes_w08} showed that
models with tapered outer edges can naturally reconcile the apparent
size discrepancy in dust and gas millimeter imaging. 
The successful fitting of CO isotopologues in the HD~163296 disk using
the SED-based disk model with an exponentially tapered outer edge, 
without invoking an unknown or unconstrained chemical effect, provides new 
support for the necessity of including this feature in outer disk structure.

The SED alone does not provide any direct information on the temperature 
structure of the intermediate disk layers where the CO (and other molecular) 
emission is generated.  In our modeling framework for the HD~163296 disk, the 
SED (and millimeter continuum images) can be fit equally well for a wide range 
of vertical temperature/density profiles, highlighting the degeneracy of the 
dust data with the parameter $z_{\rm big}$, the height marking the transition 
between the small grains in the disk atmosphere and the big grains
concentrated toward the midplane.  However, we have found that
resolved observations of optically thick CO lines at a range of
excitations can be used to place stringent constraints on the vertical
temperature structure.  Previous analysis  
of the CO J=2--1 and 3--2 lines from the HD~163296 disk suggested that gas 
temperatures were always higher than 20~K, ruling out CO freeze-out as a cause 
for the depletion of CO abundances \citep{isella_t07}.  But as we discussed in 
\S 4.2 (see Figure \ref{fig:modelspec}), the small excitation leverage between 
those low-lying transitions is not a strong discriminant of the temperature 
profile.  Here, we make use of the higher-excitation J=6--5 line to better 
constrain the vertical structure of the disk, and find a colder midplane that 
is consistent with significant CO freeze-out.  Although observations of these 
various CO transitions is expensive, there are likely other molecules
that emit at nearby frequencies and can be used to trace a sufficient
range of excitation conditions (e.g., HCO$^+$).  The essential point
is that the temperature structure in the intermediate layers of a disk
can only be constrained well using several optically thick emission
lines that probe a range of excitation.

\subsection{The ``CO Snow Line"}

Theoretical models of the molecular abundances in protoplanetary disks predict 
three distinct vertical layers \citep[see][]{bergin_a07}.  At large heights in 
the disk atmosphere, temperatures are high and molecular abundances are 
comparable to those in a standard photon-dominated region (PDR).  At 
intermediate heights, between the midplane and atmosphere, warm temperatures 
are suitable for high gas-phase abundances of many molecules.  And at the 
deepest layers near the midplane, temperatures are low enough that many 
molecules are depleted from the gas phase and frozen onto the mantles of dust 
grains.  \citet{aikawa_n06} and others, have used this kind of layered
physical  
structure in their chemical reaction network models and shown that CO is 
expected to be abundant in the warm layer, with $f_{\rm CO} \sim 10^{-4}$ 
almost independent of radius.  The boundaries of that layer are defined by the 
CO freeze-out in the midplane and the photodissociation of CO by high-energy 
stellar photons in the disk atmosphere.  The modeling of the SMA observations 
of the HD~163296 disk provides direct support for this basic model structure.  
Our estimate of the upper boundary $\sigma_s$ is in excellent
agreement with models of the upper  
PDR layer, where CO is photodissociated into its atomic constituents
by stellar UV and/or X-ray radiation at column densities
$\lesssim10^{21}$ cm$^{-2}$ \footnote{A total gas column of 10$^{21}$
  cm$^{-2}$ corresponds to 1 mag of extinction (A$_v=$1) at visible
  wavelengths when the gas-to-dust ratio is 100. But
  in our model the CO photodissociation front is located at A$_v\ll$1
  because of the large dust depletion factor as measured with
  $\epsilon=0.003$ in the upper disk layers.}   
\citep{aikawa_n06,gorti_h08}. 
The lower boundary that best fits the $^{13}$CO data, corresponding to
a temperature of 19~K, is also in excellent agreement with the
laboratory studies of CO freeze-out temperature onto dust grains
(\citealp{collings_d03,bisschop_f06}). The main effect of
CO photodesorption \citep{oberg_f07,hersant_w09} and other related
non-thermal ice desorption mechanisms is to push the lower 
boundary to colder temperatures. Since our best-fit CO lower boundary is
consistent with thermal desorption, there is no evidence for efficient
CO photodesorption in the disk of HD~163296. 

Since the term ``snow line" is often discussed in planetary formation 
as some point where the temperature in the midplane would drop below
the water ice sublimation level, e.g. in the ``minimum-mass solar nebular"
model prescribed by \citet{hayashi81}, we propose to use the term 
``CO snow line" to indicate where CO freezes out in the disk.
It was generally believed that the dust temperature in the disks of
Herbig Ae stars is high enough that even the temperature close to the
disk midplane is above the temperature of freeze-out of CO
(\citealp{dutrey_g07, jonkheid_d07}). 
In this work, the primary line of evidence for CO freeze-out in the 
disk of HD~163296 comes from the detection of sharp reduction of
the effective $^{13}$CO J=2--1 emission area in the outer disk. 
This is demonstrated in Figure~\ref{fig:zcentco} where the $^{13}$CO
visibility is best fit with the lower boundary
at 19 K, indicating no or very few $^{13}$CO emission from the
midplane area below this boundary, as shown in
Figure~\ref{fig:structure}. The CO snow 
line marks drop off at least two orders of magnitude 
in the column density of $^{13}$CO beyond 155 AU, which is clearly
resolved by our interferometric observations. However, the emission
lines from CO and its isotopologues are still detectable 
out to large radii, e.g. $>$500 AU, which indicates that there are still 
substantial amount of CO toward surface layer of the outer disk where
the temperature is higher than 19 K.

The presence of this CO snow line has a direct impact on some key
elements of chemical networks in a disk, especially related to the
ionization fraction. The CO abundance is expected to be 
linearly correlated with the HCO$^+$ abundance, but has a strong
anti-correlation with N$_2$H$^+$ since the main destruction pathway
for the latter is in reactions with CO (\citealp{jorgensen_s04}).  In
the midplane, the depletion of CO beyond the snow line should enhance
the deuterated molecular ions, H$_2$D$^+$, D$_2$H$^+$  
and D$_3^+$, such that they become the most abundant ions in the midplane 
(\citealp{ceccarelli_d05}).  So, resolving the CO snow line will
eventually  
help constrain the ionization fraction and the extent of deuterium 
fractionation in the disk interior.  

\subsection{CO Isotope Ratios}

The model developed here indicates that the gas in the intermediate layer of 
the HD~163296 disk has CO isotopologue abundance ratios that are consistent to 
those found in the molecular interstellar medium.  That finding is in contrast 
with some previous studies of this and other disks that instead suggested an 
increasing amount of depletion from $^{12}$CO to $^{13}$CO to C$^{18}$O 
\citep[e.g.,][]{isella_t07,dutrey_g94,dutrey_g96}.  That difference is
likely a  
manifestation of the underlying modeling approach, as well as the constraining 
power now available to us from the high-excitation J=6--5 line and the rare 
isotopologues. When we include CO freeze-out, our models also show no evidence 
for radial variations of the CO abundance with respect to the H$_2$ densities 
inferred (indirectly) from the dust emission.  Previous studies have found  
(much) steeper CO surface density profiles compared to dust (e.g., Pietu et 
al.~2007, Dutrey et al.~2008).  To compare with those findings, we
have used an alternative power-law prescription for the radial
abundance fractions, $f_i(r) \propto r^{p_i}$, and attempted to re-fit
our data.  The resulting best-fit values of $p_i$ for $^{13}$CO and
C$^{18}$O are $0.0\pm0.1$ and $0.1\pm0.2$, respectively: both
consistent with a radially constant abundance profile.  So, when
taking into account CO freeze-out at low temperatures, there is no
clear evidence that the radial distribution of CO deviates from the
underlying total gas surface density.  The steep slopes inferred in
previous work may be a manifestation of the sharp abundance drop
beyond the CO snow line. 

This model demonstrates that, given sufficiently strong
constraints on the vertical temperature gradient, a unified model of the
gas and dust disk can be constructed that conforms to standard ISM-based
assumptions about the isotopic abundances and the CO chemistry.

\citet{visser_v09} present a photodissociation model for CO
isotopologues including newly updated depth-dependent and isotope-selective
photodissociation rates. They find
grain growth in circumstellar disks can enhance $N(^{12}{\rm
  CO})/N({\rm C^{17}O})$ and
$N(^{12}{\rm CO})/N({\rm C^{18}O})$ ratios by a factor of ten relative to the
initial isotopic abundances. We have fit the $\sigma_s$ and
fractional abundances for C$^{17}$O and C$^{18}$O  and found that different
pairs of $\sigma_s$ and f(C$^{18}$O) or f(C$^{17}$O) can
fit the data equally well, i.e. we can not constrain $\sigma_s$  due
to the limited signal-to-noise ratios of C$^{18}$O J=2--1 and
C$^{17}$O 3--2 data. Future observations with greater sensitivity in
the CO isotopologues will be essential to investigate the
isotope-selective photodissociation on the distributions of CO
isotopologues and hence the local isotopic ratios at different layers
of the disks.  

\subsection{Comparison of HD~163296 and TW Hya}

In a series of papers, \citet{qi_h04,qi_w06} modeled the CO J=2--1, 3--2, and 
6--5 line emission from the disk around the cooler young star TW Hya.  There, 
the best fit model to the optically thick CO J=2--1 and 3--2 lines tended to 
underestimate the CO 6--5 emission.  To fit all three CO lines simultaneously 
required additional heating of the disk surface, suggested to be the result of 
the intense X-ray irradiation field for that source.  No such additional 
heating is required to explain the HD~163296 data.  On the contrary, the CO 
J=6--5 emission required a lower disk interior temperature compared to those 
predicted by the typical accretion disk models based on the SED, forcing us to 
modify the $z_{\rm big}$ parameter to fit the data. 

There are at least two plausible reasons for the different vertical structures 
we infer for these disks.  First, is the high-energy radiation field from the 
two stars. {\it Chandra} X-ray observations reveal a point-like object within
$0\farcs25$ (30 AU) of HD~163296, with an X-ray luminosity of
$4\times10^{29}$ ergs s$^{-1}$ 
(\citealp{swartz_d05}), about 5$\times$ lower than that from TW Hya
(\citealp{kastner_h99}).  Obviously the suggestion that X-ray heating is 
dependent on the stellar type will require a much larger sample to confirm, as 
well as more detailed radiative transfer modeling of the high-energy
heating of these disks.  However, it is an exciting prospect that the
interior structures of these disks may be very different depending on
the connection between the stellar mass and the irradiation
environment, with implications both for disk  
evolution and chemistry.  A second possibility may lie with the level
of turbulence in these disks.  In our model, the vertical structure is
governed by the settling of large dust grains.  The timescale for
gravitational sedimentation is expected to be short, with big grains
reaching the midplane in less than $\sim$1 Myr
(\citealp{dullemond_d04a}) unless the dust is effectively   
stirred up by turbulence.  The more vertically extended population of big 
grains required to model the HD~163296 disk might be commensurate with more 
efficient stirring, or rather inhibited settling.  \citet{hughes_w11}
suggest a 300 m s$^{-1}$ turbulent linewidth in the CO layer of the
HD~163296 disk, much larger than the $\lesssim$40 m s$^{-1}$ they find
for the TW Hya disk.  Perhaps  
turbulent stirring is responsible for lofting the big grains above the
midplane in the HD~163296 disk, enhancing the radiative cooling.

\section{Summary}

We present CO line and dust emission observations of the disk around 
the Herbig~Ae star HD~163296 and develop a model framework that describes 
in a consistent way the spectral energy distribution, resolved millimeter 
dust continuum data, and multiple emission lines of CO and its isotopologues. 
The fitting results support the general picture of CO vertical distribution 
regulated by photodissociation at a surface where the hydrogen column
density is $\lesssim10^{21}$ cm$^{-2}$ and by CO freeze-out at depths
below 20~K in the midplane. The main conclusions are summarized here:
\begin{enumerate}
\item 
We confirm a tapered exponential edge in the surface density
distribution, an outcome of the similarity solution of the time
evolution of accretion disks, can account for the size discrepancy in
dust continuum and CO emission (\citealp{hughes_w08}). This means that
constraints from dust modeling can be applied to CO gas modeling,
significantly reducing the number of free parameters. 

\item 
We find in the disk model  the transition between the ``small'' and
``big'' dust grain populations, 
$z=z_{\rm big}$ in units of the gas scale height $H$, regulates  
the vertical temperature and density profiles between the disk midplane 
and surface. Multiple transitions of CO, especially the CO 6-5 line,
which requires much higher excitation, can be used to 
constrain the location of $z_{\rm big}$, thus the temperature structure in
the disk intermediate layer. We find the resulting disk model for HD
163296 has a cold midplane populated by large grains with a large
scale height ($z_{\rm big} = 2H$). Since the vertical temperature in
the HD 163296 disk is governed mainly by the settling, or lack of
settling, of large grains, a possible explanation for the relatively
cold interior of HD 163296 disk may be a high level of gas turbulence
(see \citealp{hughes_w11}). 

\item 
Using the model with temperature structure constrained by the CO
multi-transition analysis, we fit for the location of emission from
$^{13}$CO and constrain the vertical distribution of
the emission region to lie between a lower boundary set by the
temperature where CO freezes out onto dust grains (at heights where $T
\lesssim 19$~K) and an upper boundary where densities are low enough that 
stellar and interstellar radiation can photodissociate the CO molecule
(where the vertically integrated hydrogen column density from the disk
surface is $\lesssim10^{21}$ cm$^{-2}$).
The CO freeze-out produces a significant drop in the gas-phase CO column 
density beyond a radius of about 155 AU, effectively a CO snow line that 
is resolved directly by the observations. The CO depletion, generally
found in disks, can be successfully accounted for considering both the
CO freeze-out and photodissociation.   

\item 
Taking the CO freeze-out into consideration in the disk model and
assuming CO isotopologues sharing the same spatial distribution, the
isotopic ratios of $^{12}$C/$^{13}$C, $^{16}$O/$^{18}$O and 
$^{18}$O/$^{17}$O are consistent with the standard quiescent interstellar 
gas-phase values and show no evidence for unusual
fractionation. More sensitive data is needed to investigate the
distribution differences among those CO isotopologues and the local
isotopic ratios at different layers of the disks. 

\end{enumerate}

This detailed investigation of the HD~163296 disk demonstrates the 
potential of a staged, parametric technique for constructing unified
gas and dust structure models and constraining the distribution of 
molecular abundances using resolved multi-transition, multi-isotope
observations. The analysis provides the essential framework for more
general observational studies of molecular line emission in
protoplanetary disks.   

We thank Edwin Bergin, Eugene Chiang and Jeremy Drake for beneficial
conversations, and a referee for constructive comments on the paper.
Support for K.~I.~O. is provided by NASA through a Hubble Fellowship
grant awarded by the Space Telescope Science Institute, which is
operated by the Association of Universities for Research in Astronomy,
Inc., for NASA, under contract NAS 5-26555. Support for A.~M.~H. is
provided by a fellowship from the Miller Institute for Basic Research
in Science. We acknowledge NASA Origins of Solar Systems grant
No. NNX11AK63G.    

\bibliographystyle{apj}

\clearpage

\begin{deluxetable}{lcccccc}
\tabletypesize{\scriptsize}
\rotate
\tablewidth{0pt}
\tablecaption{Observational parameters for SMA HD~163296 observations\label{tab:obs1}}
\tablehead{
\colhead{Parameters}&\colhead{CO 2--1}&\colhead{CO 3--2$\tablenotemark{b}$}& 
\colhead{CO 6--5} & \colhead{$^{13}$CO 2--1} &
\colhead{C$^{18}$O 2--1}  &\colhead{C$^{17}$O 3--2}}
\startdata
Rest Frequency (GHz): & 230.53800 & 345.79599 & 
691.47308 & 220.39868 & 
219.56036 & 337.06113 \\
Observations$\tablenotemark{a}$   & 2010 May 17 (C) & 2009 May 6 (C) & 
2007 Mar 20 (C) & 2010 May 15 (C) & 
2010 May 15 (C)  & 2005 Aug 23 (C) \\
& 2010 Sep 14 (E) & 2009 Aug 22(E) & 
& 2010 Sep 11 (E) & 
2010 Sep 11 (E)   & \\
Tsys & 93--207 & 147--273 &
1368--4764 & 82--146 & 
82--146 & 210--603 \\
Baselines & 6--151 & 10--146 &
23--158 & 6--145 &
6--145 & 8--80 \\
Spectral resolution (km s$^{-1}$): & 0.066 & 0.044 & 0.88 & 0.069 &
0.55 & 0.72 \\
\enddata
\tablenotetext{a} { C: compact configuration; E: extended configuration. }
\tablenotetext{b} { Observational parameters from ~\citet{hughes_w11}}
\end{deluxetable}

\begin{deluxetable}{lccc}
\tablewidth{0pt}
\tablecaption{Continuum and emission line results \label{tab:cont}}
\tablehead{
\colhead{$\lambda$(mm)}&\colhead{Beam}&\colhead{PA}&\colhead{Flux (Jy)}}
\startdata
1.36 & $3\farcs5 \times 2\farcs2$ & 55.8$^\circ$ & 0.615$\pm$0.004 \\ 
1.33 & $4\farcs7 \times 2\farcs8$ & 24.2$^\circ$ & 0.670$\pm$0.007 \\
1.11 & $3\farcs0 \times 1\farcs9$ & 57.8$^\circ$ & 1.039$\pm$0.007 \\
0.88 & $1\farcs8 \times 1\farcs3$ & 47.6$^\circ$ & 1.74$\pm$0.12 \\
0.44 & $2\farcs7 \times 2\farcs2$ & -6.9$^\circ$ & 7.5$\pm$0.5 \\
\hline
\hline
Transitions & Beam & PA & Integrated Intensity$\tablenotemark{b}$ (Jy
km$^{-1}$) \\
\hline
CO 2--1 & $2\farcs1 \times 1\farcs8$ & 10.5$^{\circ}$ & 54.17$\pm$0.39\\
CO 3--2 & $1\farcs7 \times 1\farcs3$ & 46.7$^{\circ}$ &
98.72$\pm$1.69\\
CO 6--5 & $2\farcs7 \times 1\farcs4$ & 1.8$^{\circ}$  &
58.66$\pm$6.44\\
$^{13}$CO 2--1 & $1\farcs9 \times 1\farcs8$ & 5.5$^{\circ}$ &
18.76$\pm$0.24\\
C$^{18}$O 2--1 & $1\farcs9 \times 1\farcs8$ & 5.5$^{\circ}$ &
6.30$\pm$0.16\\
C$^{17}$O 3--2 & $3\farcs2 \times 2\farcs2$ & 13.1$^{\circ}$ &
11.64$\pm$0.76\\ 
\enddata
\tablenotetext{a} { Intensity averaged over the whole emission area.}
\end{deluxetable}

\begin{deluxetable}{l c}
\tablewidth{0pt}
\tablecaption{Physical model for the disk of HD 163296 \label{tab:model}}
\tablehead{
Parameters & Values }
\startdata
\multicolumn{2}{c}{Stellar and accretion properties} \\
\hline
Spectral type & A1 \\
Effective temperature: T$_*$(K) & 9333\\
Visual extinction: A$_v$ & 0.3 \\
Estimated distance: d(pc) & 122\\
Stellar radius: R$_*$(R$_\odot$) & 2\\
Stellar mass: M$_*$(M$_\odot$)& 2.3 \\
Accretion rate: $\dot{M}$($M_{\odot}$ yr$^{-1}$) &
7.6$\times$10$^{-8}$\\
\hline
\multicolumn{2}{c}{Disk structure properties} \\
\hline
Disk mass: M$_d$(M$_\odot$) & 0.089 \\
Characteristic radius: R$_c$(AU) & 150 \\
Viscosity coefficient: $\alpha_0$ & 0.019 \\
Depletion factor of the atmospheric small grains: $\epsilon^a$ & 0.003 \\
$z_{\rm big}$$^a$ ($H^b$) & 2.0 \\ 
Inner wall radius: R$_{wall}$(AU) & 0.6 \\
Inner wall scale height: z$_{wall}$(AU) & 0.1 \\
\hline
\multicolumn{2}{c}{Disk geometric and kinematic properties} \\ 
\hline
Inclination: $i$(deg) & 44 $\pm$ 2 \\
Systemic velocity: V$_{LSR}$(km s$^{-1}$) & 5.8 $\pm$ 0.2 \\
Turbulent line width$^c$: v$_{turb}$(km s$^{-1}$) &0.2 \\
Position angle: P.A.(deg) & 133 $\pm$ 3 \\
\enddata
\tablenotetext{a}{See definition in paper or \citet{dalessio_c06}.}
\tablenotetext{b}{Gas scale height.}
\tablenotetext{c}{Fixed parameter.}
\end{deluxetable}

\begin{deluxetable}{lcccc}
\rotate
\tablewidth{0pt}
\tablecaption{Fitting results: fractional abundances and distributions \label{tab:model2}}
\tablehead{
\colhead{Parameters}&\colhead{CO 2--1 \& 3--2}&\colhead{$^{13}$CO 2--1}&
\colhead{C$^{18}$O 2--1} &\colhead{C$^{17}$O 3--2} }
\startdata
Midplane freeze-out temperature (K) & 19.0$^*$ & 19.0 $\pm$ 0.3 & 19.0$^*$& 19.0$^*$\\
$\sigma_s$ & 0.79$^*$ & 0.79 $\pm$ 0.03 & 0.79$^*$ & 0.79$^*$ \\ 
Fractional abundance  & (6.0$\pm$0.3) $\times$10$^{-5}$ & (9.0$\pm$0.6)
$\times$10$^{-7}$ & (1.35$\pm$0.20) $\times$10$^{-7}$ & (3.5$\pm$1.1)
$\times$10$^{-8}$ \\
Fractional abundance (no freeze-out) & (6.5$\pm$0.3) $\times$10$^{-5}$ &
(4.6$\pm$0.3) $\times$10$^{-7}$ & (1.5$\pm$0.2) $\times$10$^{-8}$ &
(7.0$\pm$2.2) $\times$10$^{-8}$ \\
\enddata
\tablenotetext{*} { Parameter values adopted from $^{13}$CO 2--1 fitting. }
\end{deluxetable}

\clearpage

\clearpage

\begin{figure}[t]
\epsscale{1.0}
\plotone{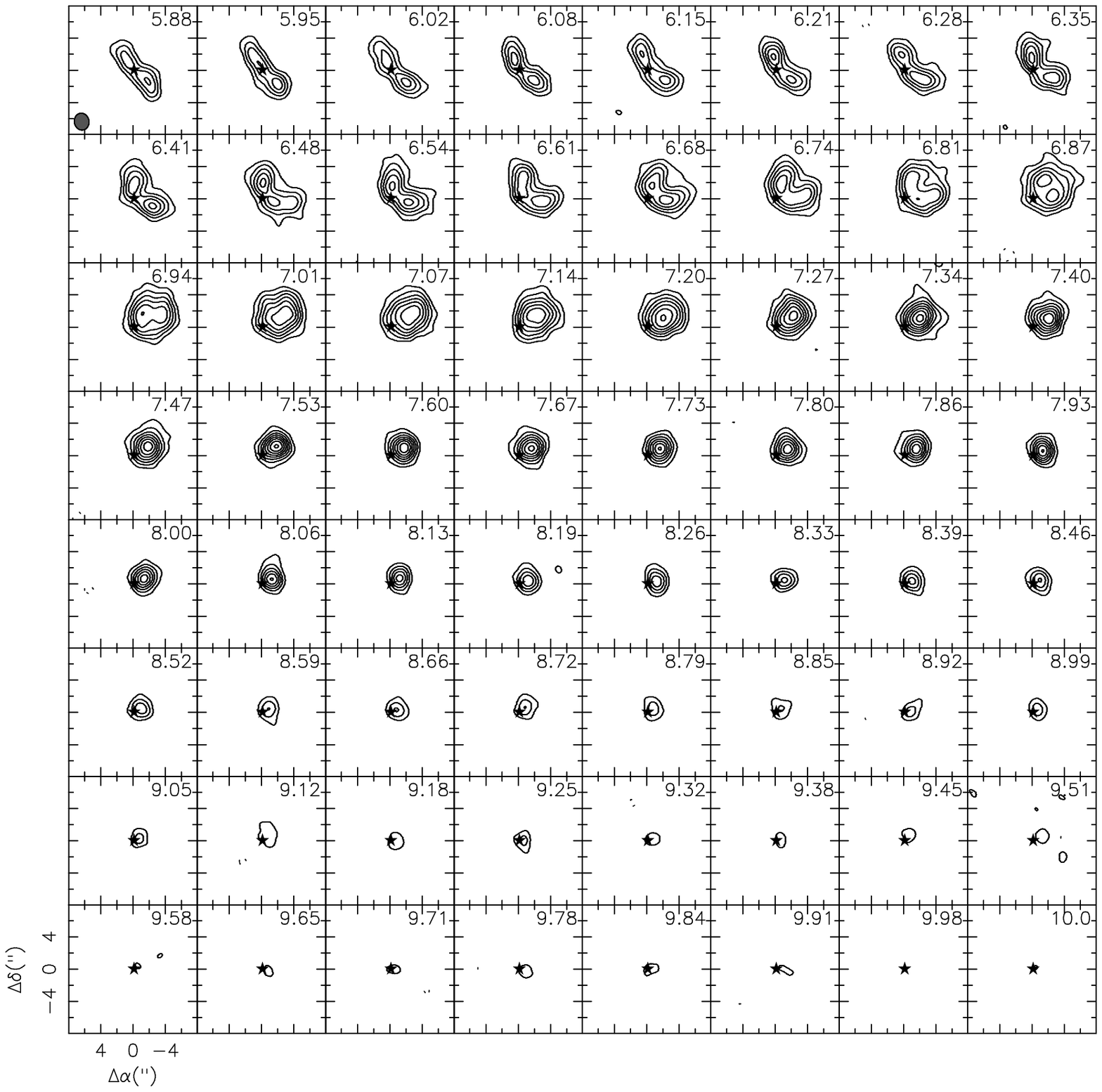}
\end{figure}

\clearpage

\begin{figure}[t]
\epsscale{1.0}
\plotone{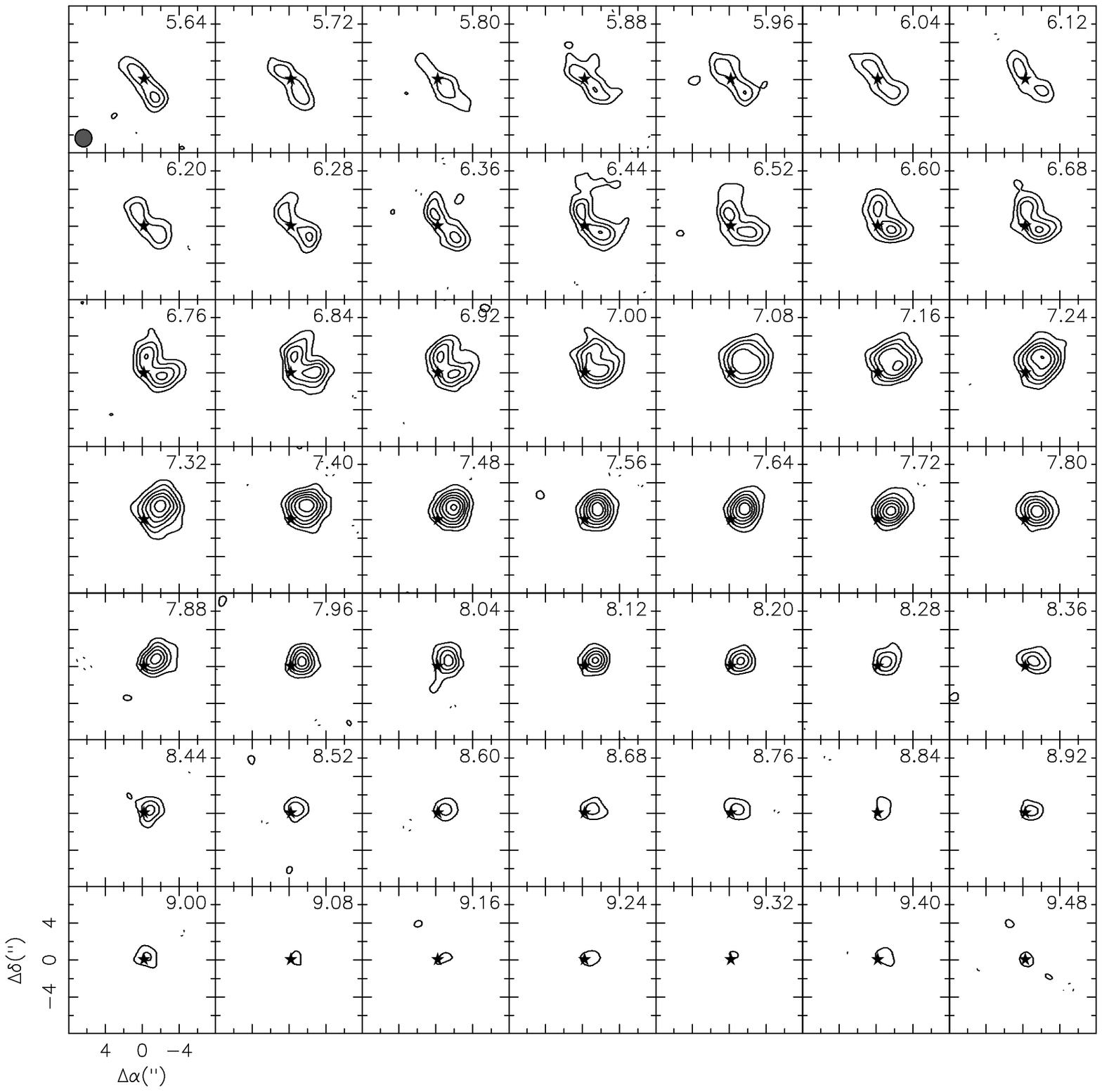}
\end{figure}

\clearpage

\clearpage

\clearpage

\clearpage

\clearpage

\clearpage

\clearpage

\clearpage

\clearpage

\clearpage

\end{document}